\documentclass[apj]{emulateapj}

\usepackage{times}
\usepackage{natbib}
\usepackage{amsmath}
\usepackage{graphicx}

\newcommand{\str}{Str\"{o}mgren }
\newcommand{\ars}{\langle R_s \rangle}
\newcommand{\nH}{n_{\rm H, \infty}}

\newcommand{\nfive}{10^5~{\rm cm^{-3}}}
\newcommand{\nfour}{10^4~{\rm cm^{-3}}}
\newcommand{\nthree}{10^3~{\rm cm^{-3}}}
\newcommand{\ntwo}{10^2~{\rm cm^{-3}}}
\newcommand{\lambdaavg}{\langle \lambda_{\rm rad} \rangle}

\newcommand{\hi}{\rm H~{\textsc i}}
\newcommand{\hii}{\rm H~{\textsc {ii}}}

\newcommand{\mach}{\mathcal{M}}

\newcommand{\myscale}{1.1}
\newcommand{\vbh}{v_{\rm bh}}

\shorttitle{Radiation-Regulated Accretion onto Black Holes in Motion}
\shortauthors{Park \& Ricotti}

\received{}
\accepted{}
\begin{document}

\setlength{\pdfpageheight}{\paperheight}
\setlength{\pdfpagewidth}{\paperwidth}

\title{ACCRETION ONTO BLACK HOLES FROM LARGE SCALES REGULATED BY RADIATIVE FEEDBACK. III. ENHANCED LUMINOSITY OF INTERMEDIATE MASS BLACK HOLES MOVING AT SUPERSONIC SPEEDS}

\author{KwangHo Park\altaffilmark{1} and Massimo Ricotti \altaffilmark{2}} 
\affil{Department of Astronomy, University of Maryland, College Park, MD 20740, USA} 
\email{kpark@astro.umd.edu, ricotti@astro.umd.edu}

\altaffiltext{1}{McWilliams Center for Cosmology, Carnegie Mellon University, Pittsburgh, PA 15213, USA}
\altaffiltext{2}{Joint Space-Science Institute (JSI), College Park, MD 20742, USA}

\begin{abstract}
In this third paper of a series, we study the growth and luminosity
of black holes (BHs) in motion with respect to their surrounding
medium. We run a large set of two-dimensional axis-symmetric
simulations to explore a large parameter space of initial conditions
and formulate an analytical model for the accretion. Contrary to the
case without radiation feedback, the accretion rate increases with
increasing BH velocity $\vbh$ reaching a maximum value at $\vbh=2c_{\rm
s,in}\sim 50~{\rm km}~{\rm s^{-1}}$, where $c_{\rm s,in}$ is the sound speed inside the
``cometary-shaped'' \hii~region around the BH, before decreasing as
$\vbh^{-3}$ when the ionization front (I-front) 
becomes {\it R}-type (rarefied) and
the accretion rate approaches the classical Bondi--Hoyle--Lyttleton
solution. The increase of the accretion rate with $\vbh$ is produced
by the formation of a {\it D}-type (dense) I-front
preceded by a standing bow-shock that reduces the downstream gas velocity
to transonic values. There is a range of densities and velocities where
the dense shell is unstable producing periodic accretion rate peaks which
can significantly increase the detectability of intermediate-mass BHs.
We find that the mean accretion rate for a moving BH is larger than
that of a stationary BH of the same mass if the medium temperature is
$T_\infty<10^4$~K. This result could be important for the growth of seed
BHs in the multi-phase medium of the first galaxies and for building
and early X-ray background that may affect the formation of the first
galaxies and the reionization process.

\end{abstract}



\section{Introduction} 
\label{sec:intro}

Gravitationally driven gas inflow onto moving point masses such as black
holes (BHs) or neutron stars has been described analytically in the 1940s
by Bondi--Hoyle--Lyttleton \citep*{HoyleL:39,BondiH:44,Bondi:52}. The
generalized formula for the accretion onto a point mass moving with
velocity $\vbh$ is obtained from the Bondi formula by replacing the gas
sound speed, $c_{\rm s, \infty}$, with an effective speed $v_{\rm
  eff}=(c_{\rm s,\infty}^2+\vbh^2)^{1/2}$. The accretion rate $\dot{M}
\propto \rho v_{\rm eff}\sigma_{\rm eff}$ is roughly estimated as the
gas flux through the effective cross section $\sigma_{\rm eff}=\pi
r_{\rm eff}^2$, where $r_{\rm eff}=GM_{\rm bh}/v_{\rm eff}^2$ is the distance
at which the escape velocity of the gas equals $v_{\rm eff}$. 
Despite the similarities between the Bondi and Lyttleton
formulae, in the second case the accretion onto the BH is not
spherically symmetric: most of the gas streams past the BH and is
gravitationally focused on the axis of symmetry of the problem. The
component of the gas velocity perpendicular to the direction of the BH
motion is converted into thermal energy and is mostly dissipated, thus
a fraction of the gas becomes gravitationally bound to the BH and is
accreted from the downstream direction.  The generalized
Bondi--Hoyle--Lyttleton formula is
\begin{equation}
\dot{M}_{\rm BHL} = \frac{\dot{M}_B} {(1+\vbh^2/c_{\rm s,\infty}^2)^{3/2}}, 
\label{eq:bhl}
\end{equation}
where $\dot{M}_B$ is the Bondi accretion rate onto non-moving BHs
$\dot{M}_B = \pi e^{3/2} \rho_\infty G^2M_{\rm bh}^2
c_{\rm s,\infty}^{-3}$ (assuming isothermal equation of state: $\gamma
=1$). Here, $M_{\rm bh}$ is the BH mass, and $\rho_{\infty}$ is the
density of the ambient gas. The term in the denominator of
Equation~(\ref{eq:bhl}) is the only term that accounts for the motion
of the point mass, thus for supersonic BH motion the accretion rate
decreases with increasing velocity as $\vbh^{-3}$. Numerical simulations
of accretion onto moving BHs have confirmed the validity of the
Lyttleton equation \citep{Shima:85,Ruffert:94,Ruffert:96}, even in the
presence of non-axisymmetric flows that necessarily arise in 
three-dimensional 
simulations due to hydrodynamical instabilities
\citep{Cowie:77,MatsudaIS:87,FryxellT:88,TaamF:88,Soker:90,
  KoideMS:91,LivioSMA:91,FoglizzoR:97,FoglizzoR:99,Foglizzo:05}.

For the case of accretion onto stationary BHs, the radiation emitted
near the gravitational radius of the BH typically reduces the rate of
gas supply to the BH from large scales well below the value given by
the Bondi formula.  This is because X-ray and UV heating increases the
sound speed of the gas in the proximity of the BH
\citep*{OstrikerWYM:76,CowieOS:78,BB:80,
  KrolikL:83,Vitello:84,WandelYM:84,OstrikerCCNP:10}, and because
radiation pressure accelerates the gas and dust away from the BH
\citep*{shapiro:73, OstrikerWYM:76,Begelman:85, Ricotti:08}.  Indeed,
the maximum accretion rate that can be achieved in most cases is the
Eddington rate. In this limit, the outward acceleration of the gas due
to Compton scattering of radiation with free electrons equals the
inward gravitational acceleration.  For supermassive BHs (SMBHs), the
picture is far more complex because the hot ionized region and the
Bondi radius extend to galactic scales. Thus, it is necessary to model
the evolution of the host galaxy, with its stellar populations and the
interstellar medium (ISM) in addition to the accreting SMBH.  Despite the
complexity of this problem, a significant amount of work exists in the
literature on the self-regulation mechanisms for the growth of SMBHs
at the centers of elliptical galaxies \citep*{CiottiO:01,
  Sazonov:05,CiottiO:07,CiottiOP:09,LussoC:11, NovakOC:11,NovakOC:12},
for radiation-driven axis-symmetric outflows in active galactic nuclei
\citep*{Proga:07,ProgaOK:08,KurosawaPN:09,KurosawaP:09a,KurosawaP:09b}
and more generally on the co-evolution of galaxies and their SMBHs. The
study of the cosmological origin and growth of SMBHs is a multi-scale
problem often approached using large-scale simulations. Typically, the
limited resolution of these cosmological simulations precludes
resolving the SMBH Bondi radius, hence, the accretion rate is modeled
using sub-grid recipes largely based on the Eddington-limited Bondi
formula
\citep*{Volonteri:05,Pelupessy:07,Greif:08,AlvarezWA:09,KimWAA:11,BlechaCLH:11,BlechaLN:12},
sometimes with the introduction of a normalization parameter for the
accretion rate to match observations
\citep{DiMatteo:05,SpringelDH:05,DiMatteo:08}.

Recently, there has been renewed interest on studies of accretion onto
intermediate mass BHs 
\citep[IMBHs;][]{MiloBCO:09,MiloCB:09,ParkR:11,Li:11,ParkR:12a}, motivated by
observations of ultraluminous X-ray sources (ULXs) in nearby galaxies
\citep*[for a review][]{Miller:04,vanderMarel:04}, and in the early
universe, because some simulations predict formation of IMBHs as
Population~III star remnants
\citep*{AbelANZ:98,BrommCL:99,AbelBN:00,MadauR:01,SchneiderFNO:02,OhH:02,
  WhalenF:12,Jeonetal:11,JohnsonWFL:12,StacyGB:12} or from direct
collapse of quasi-stars
\citep*{Carr:84,HaehneltNR:98,Fryer:01,BegelmanVR:06,
  VolonteriLN:08,OmukaiSH:08,ReganH:09,MayerKEC:10,JohnsonKGD:11,JohnsonWLH:12}.

Our previous work in this series \citep[][hereafter Papers~I and II,
  respectively]{ParkR:11, ParkR:12a} has been partially motivated by
the need to better describe the sub-grid recipe for accretion onto
IMBH (and SMBH, with some caveats) in large-scale cosmological
simulations. We have taken the most basic approach to understanding
the growth of IMBHs and SMBHs, focusing on the fundamental problem of
Bondi-type accretion (accretion from a homogeneous and isotropic gas)
modified by a simple feedback loop that couples the accretion rate to
the luminosity output: $L=\eta {\dot M}c^2$. The photon--gas coupling
is due to isotropic emission of radiation from a central source,
affecting the gas inflow through thermal and radiation pressures. We
have started with idealized initial conditions, adding one physical
process at a time, in order to understand and model analytically the
modified Bondi problem for a large parameter space of initial
conditions. One-dimensional and two-dimensional (2D) radiation-hydrodynamic simulations are used to
explore the large parameter space of BH masses, ambient gas
density/temperature, radiative efficiency, and spectrum of the emitted
radiation. We have then derived a physically motivated model for the
BH growth and luminosity and provide analytic formulae that fully
characterize the model. Despite the simplicity of the initial
conditions, we found a rich phenomenology that can be described with a
simple scaling arguments.  The main results are: the mean accretion
rate is proportional to the thermal pressure of the ambient gas and is
always less than 1\% of the Bondi rate. The accretion rate is
periodic, with period of the luminosity bursts proportional to the
average size of the ionized hot bubble and peak accretion rates about
10 times the mean. Most interestingly, we have discovered that there
are two distinct modes of oscillations with very different duty cycles
(6\% and 50\%), governed by two different depletion processes of the
gas inside the ionized bubble (see Papers~I and II).

This paper is the continuation of our previous work on characterizing how
the Bondi--Lyttleton problem is modified by radiation feedback, but here
we focus on the growth rate and luminosity of BHs in motion with respect
to their surrounding medium: i.e., Lyttleton accretion modified by a
radiative feedback loop. Considering the motion of the BH is important for
the study of stellar BH \citep[e.g.,][]{WheelerJ:11} and IMBHs accreting
from the ISM, both in the early universe and in the local galaxies as
candidates for ULXs \citep*{Krolik:81,Krolik:84,Krolik:04,RicottiO:04b,
RicottiOG:05,Ricotti:07,StrohmayerM:09}, and SMBHs in merging
galaxies. Both radiation feedback and BH motion are expected to reduce
the accretion rate with respect to the Bondi rate, but it is unknown
what is the combined effect of these two processes. To the best of our
knowledge, the present study is the first to consider radiation feedback
effects on moving BHs.

This paper is organized as follows. In Section~\ref{sec:numerical},
we introduce basic definitions used throughout this series of papers
and describe the numerical simulations. In Sections~\ref{sec:results}
and~\ref{sec:modelling}, we show our simulation results and the analytical
model describing the set of simulations respectively. Finally,
we summarize and discuss the implications of our work in
Section~\ref{sec:discussion}.

\begin{figure}[t] 
\epsscale{\myscale} 
\plotone{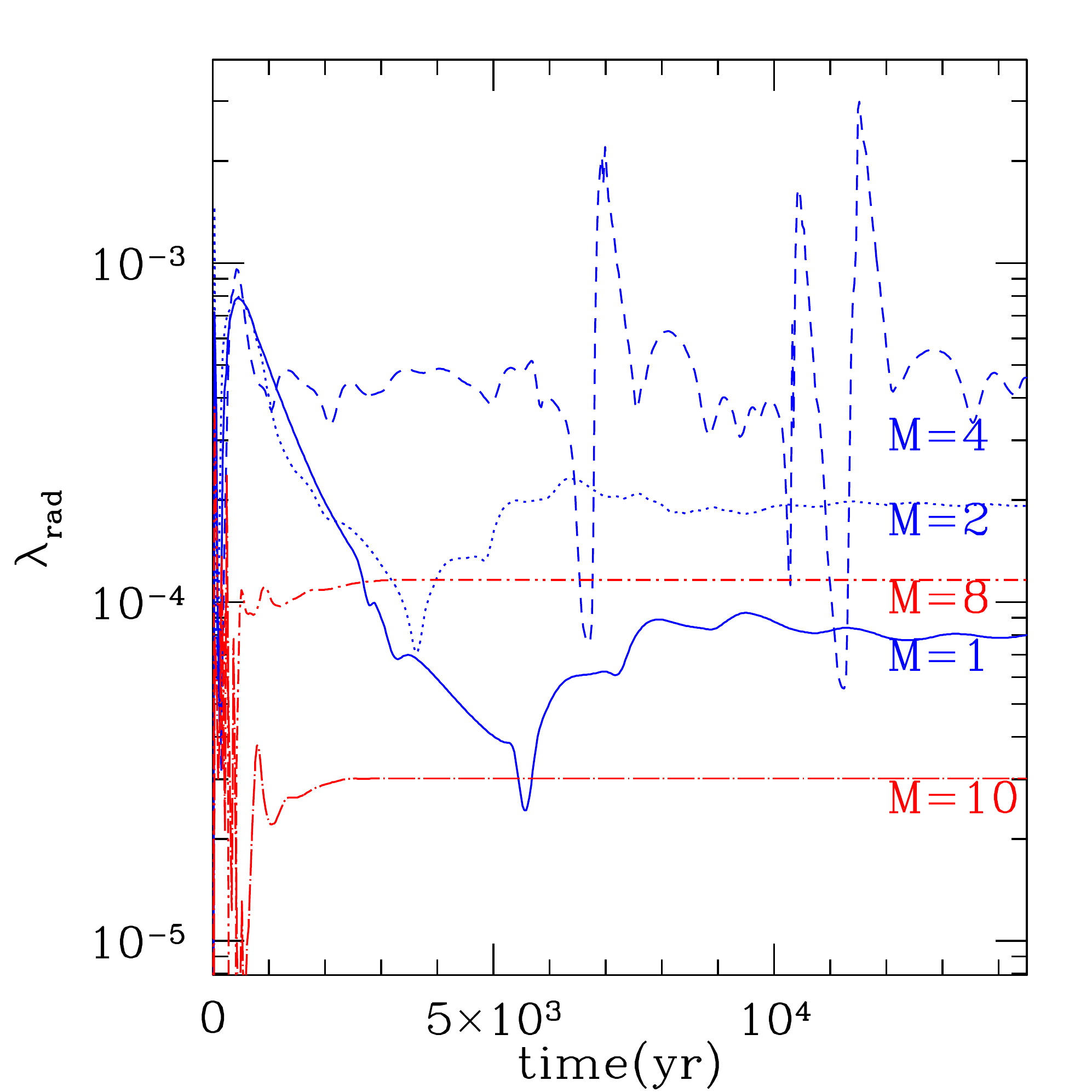}
\caption
{Accretion rate in units of the Bondi rate as a function
  of time for simulations with $M_{\rm bh} = 100~M_\odot$,
  $\nH=\nthree$, and $T_\infty=10^4~{\rm K}$. The lines show
  simulations for BH moving at Mach numbers $\mach=1$, 2, 4, 8, and 10 (see
  labels under each line). The early phase in all the simulations
  shows oscillatory behavior of the accretion rate due to
  non-equilibrium initial conditions. However, in most simulations the
  accretion rate quickly reaches a steady state. The average accretion
  rate increases with increasing Mach number for Mach numbers from 1
  to 4 and decreases for larger Mach numbers. Quasi-periodic bursts of
  accretion are seen in the simulation with $\mach= 4$.}
\label{acc_d3} 
\end{figure}
\section{Basic Definitions and Radiation-Hydrodynamic Simulations}
\label{sec:numerical}

For the sake of consistency with the previous papers of this series, we
define the dimensionless accretion rate $\lambda_{\rm rad} \equiv
\dot{M}/\dot{M}_B$, where $\dot{M}_B$ is the Bondi
accretion rate for isothermal gas ($\gamma =1$). In Paper~II we found
that the mean accretion rate for a stationary BHs regulated by
radiation feedback is
\begin{equation}
\langle \dot{M} \rangle = {\rm min}~(1\% T_{\infty, 4}^{5/2} 
\dot{M}_{B}, \eta^{-1} \dot{M}_{\rm Edd}), 
\label{eq:1percent}
\end{equation}
where we have defined $T_{\infty, n} \equiv T_\infty/(10^n~{\rm
  K})$. Equation~(\ref{eq:1percent}) is valid for density $n_{\rm H,
  \infty} \gtrsim 10^5 M_2^{-1}~{\rm cm^{-3}}$ where a similar
definition $M_n \equiv M_{\rm bh}/ (10^n~M_\odot)$ is made. Instead,
if $n_{\rm H, \infty} \lesssim 10^5 M_2^{-1}~{\rm cm^{-3}}$, the
dimensionless accretion rate in the sub-Eddington regime shows a
dependency with the density $\lambdaavg = 1\% T_{\infty,4}^{5/2}
n_{5}^{1/2}$, where again we define $n_{m} \equiv \nH / 10^m~{\rm
  cm}^{-3}$. The Eddington luminosity and accretion rates are defined
as $L_{\rm Edd}= 4 \pi G M_{\rm bh} m_p c \sigma_{T}^{-1}= (3.3 \times
10^6~L_{\sun}) M_2$, and $\dot M_{\rm Edd} \equiv L_{\rm Edd} c^{-2}$,
respectively. Note that we have adopted a definition of $\dot M_{\rm
  Edd}$ independent of the radiative efficiency $\eta$, thus for
Eddington limited accretion the accretion rate in units of $\dot
M_{\rm Edd}$ is $\dot m = 1/\eta>1$.

The numerical method used in this paper is the same as in Papers~I and
II. We run a set of 2D radiation-hydrodynamic simulations using a
modified parallel version of the non-relativistic hydrodynamics code
ZEUS-MP \citep{StoneN:92,Hayes:06} with our photon-conserving UV and
X-ray one-dimensional radiative transfer equation solver
\citep{RicottiGS:01,WhalenN:06}. We include photo heating,
photo ionization, and chemistry of H/He in multi-frequency are
manifested in the code to see how high-energy UV and X-ray photons
regulate gas accretion onto moving BHs. Radiation pressure both on
electrons and \hi~is calculated to simulate the effect of momentum
transfer from ionizing photons to gas. See Paper~I and Paper~II for
detailed description.

Axis-symmetric geometry with respect to the azimuthal angle ($\phi$)
is applied to all simulations.  We use logarithmically spaced grid in
the radial direction ($r$), and evenly spaced grid in the polar angle
direction ($0 \le \theta \le \pi $), with BH centered at the origin
$r=0$. Axis-symmetric configuration is necessary to simulate BHs in
motion relative to ambient gas which is assumed to be moving parallel
to the polar axes. In the radial direction we apply flow-in boundary
conditions at the outer boundary for the first half of the polar angle
($0 \le \theta \le 0.5 \pi$) and flow-out boundary conditions for the
second half ($ 0.5\pi < \theta \le \pi$). At the inner boundary, in
the radial direction, we apply flow-out boundary condition for the
entire polar angle. Reflective boundary conditions are applied along
the polar axis $\theta=0$ and $\theta=\pi$ to satisfy the
axis-symmetric configuration.

\begin{figure}[t] 
\epsscale{\myscale} 
\plotone{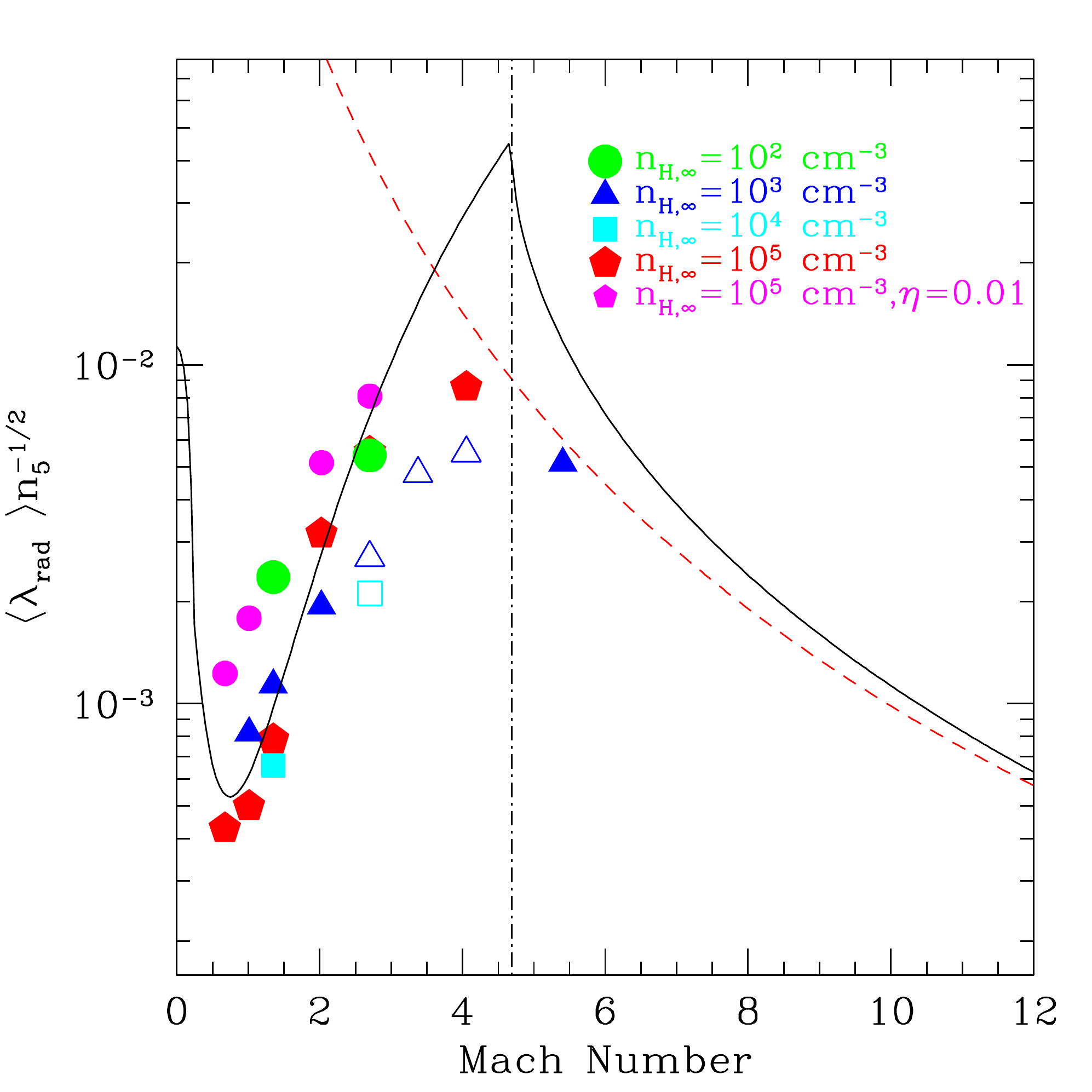}
\caption{Mean accretion rate in units of the Bondi rate $\lambdaavg$
  as a function of BH Mach number. Classical Hoyle--Lyttleton accretion
  predicts a monotonic decrease of accretion rate as $\lambdaavg
  \propto (1+\mach^2)^{-3/2}$ (dashed line). However our simulations
  (solid and open symbols) show that $\lambdaavg$ decreases with
  $\mach$ for subsonic velocities, has a minimum at $\mach \sim 1$,
  increases for $1< \mach \lesssim \mach_R$, and decreases again for
  $\mach > \mach_R$.  The solid line shows our model based on {\it
    D}-type I-front jump conditions preceded by an isothermal
  bow-shock forming in the upstream direction for $\mach < \mach_R$
  and a transition to {\it R}-type I-front for Mach numbers larger
  than $\mach_R$ (shown by the dot-dashed line). The $Y$-axis shows
  $\lambdaavg/\nH^{1/2}$ because simulations with ambient densities
  $\nH<\nfive$ have $\lambdaavg \propto \nH^{1/2}$, in agreement with
  previous results for non-moving BHs (see Paper~I). Solid symbols
  show simulations that reach a steady-state solution, while open
  symbols indicate simulations that present oscillatory behavior of
  the accretion rate due to instabilities of the thin-shell downstream
  of the bow shock.}
\label{lambda}
\end{figure}

We assume uniform density and constant velocity for the initial
conditions. For the supersonic cases ($\mach > 1$), we start the
simulations with an assumption of fixed accretion rate ($\lambdaavg =
0.001$) to reduce the effect of oscillation observed at early phases
of the simulations. For our reference simulations, we select a fiducial
value of the radiative efficiency $\eta=0.1$, typical for thin disk
models \citep{ShakuraS:73}, BH mass $M_{\rm bh}=100~M_\odot$, and the
temperature of the ambient gas $T_\infty=10^4~{\rm K}$. We explore a
range of Mach numbers up to $\mach =10$, and gas densities
$\nH=10^2$--$\nfive$.

\begin{figure*}[th]
\epsscale{\myscale} 
\plotone{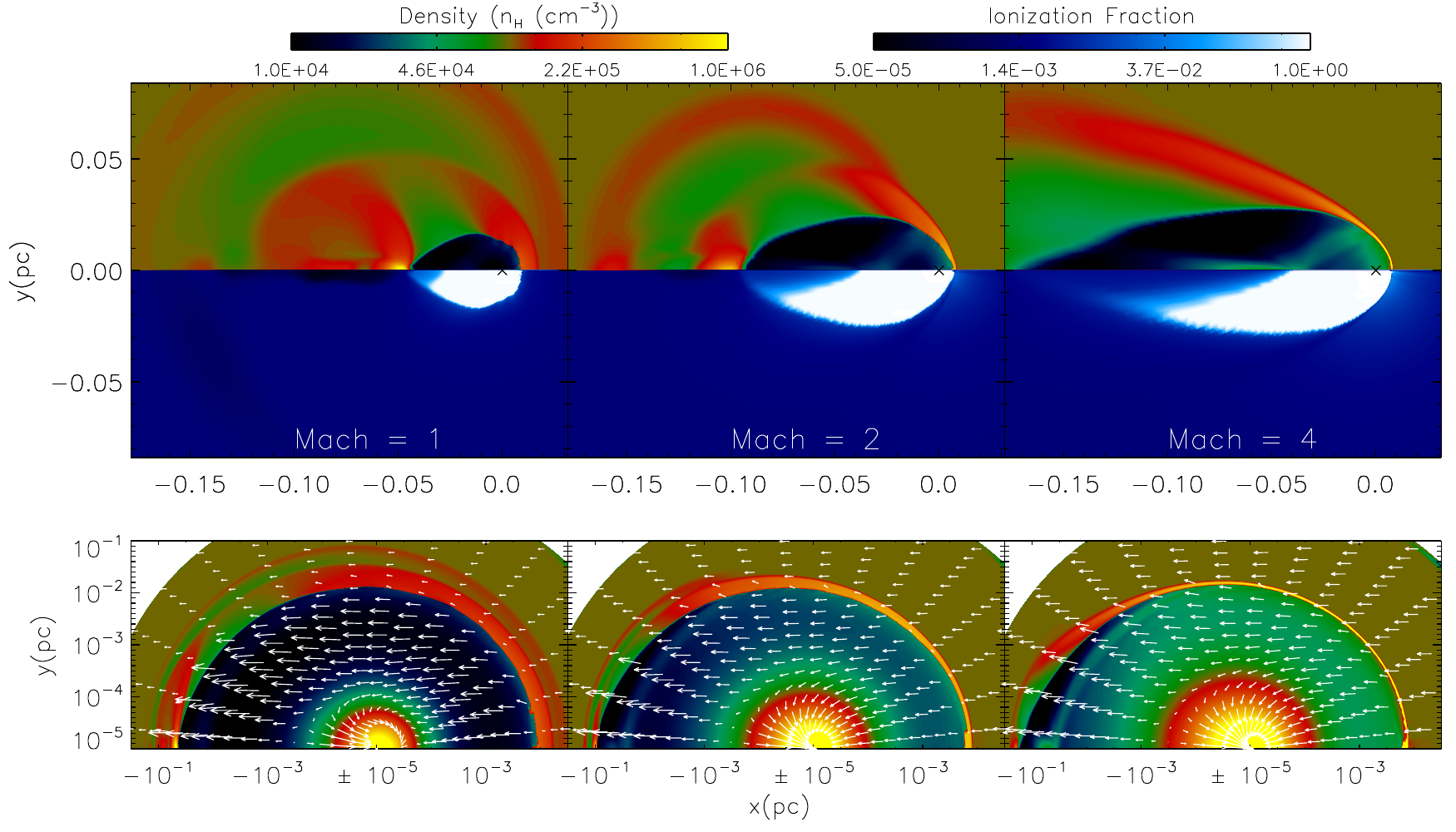}
\caption{{Top:} density (top half of the panel) and ionization fraction
  (bottom half of the panel) for simulations of a BH of mass $M_{\rm
    bh} = 100~M_\odot$, gas density $\nH=\nfive$, and temperature
  $T_{\infty}=10^4~$K moving at $\mach=1,2, {\rm and}~4$ (from left to
  right).  Each panel shows a large-scale view of the cometary-shaped
  \hii~region and the dense shell in the upstream direction for
  simulation with different Mach numbers. The size of the \str sphere
  in the downstream direction increases roughly linearly with
  increasing Mach number but remains roughly constant as a function of
  Mach number in the upstream direction. With increasing Mach number,
  the density of the shell in the upstream direction increases
  ($(n_{\rm H, sh}/n_{\rm H, \infty}) \propto \mach^2$) and the
  density behind the dense shell also increases $n_{\rm H,in}/n_{\rm
    H, \infty} \propto \mach^2$ for $1< \mach < \mach_R$. {Bottom:} the
  same simulations as in the top panel but showing the gas density field and
  gas velocity vectors adopting logarithmic scale in the radial
  direction to emphasize the gas flow at small scales.}
\label{den_d5}
\end{figure*}

\section{Numerical Results}
\label{sec:results}

\subsection{Accretion Rate as a Function of Mach Number}
\label{ssec:lambda_mach}

The classical Bondi--Hoyle--Lyttleton accretion predicts a monotonic decrease
of accretion rate with increasing velocity of the BH with respect to
the ambient gas as in Equation~(\ref{eq:bhl}). However, our simulations
of moving BHs with radiative feedback show a very different dependence
of accretion rate as a function of the Mach number.

\begin{figure*}[th]
\epsscale{1.0} 
\plotone{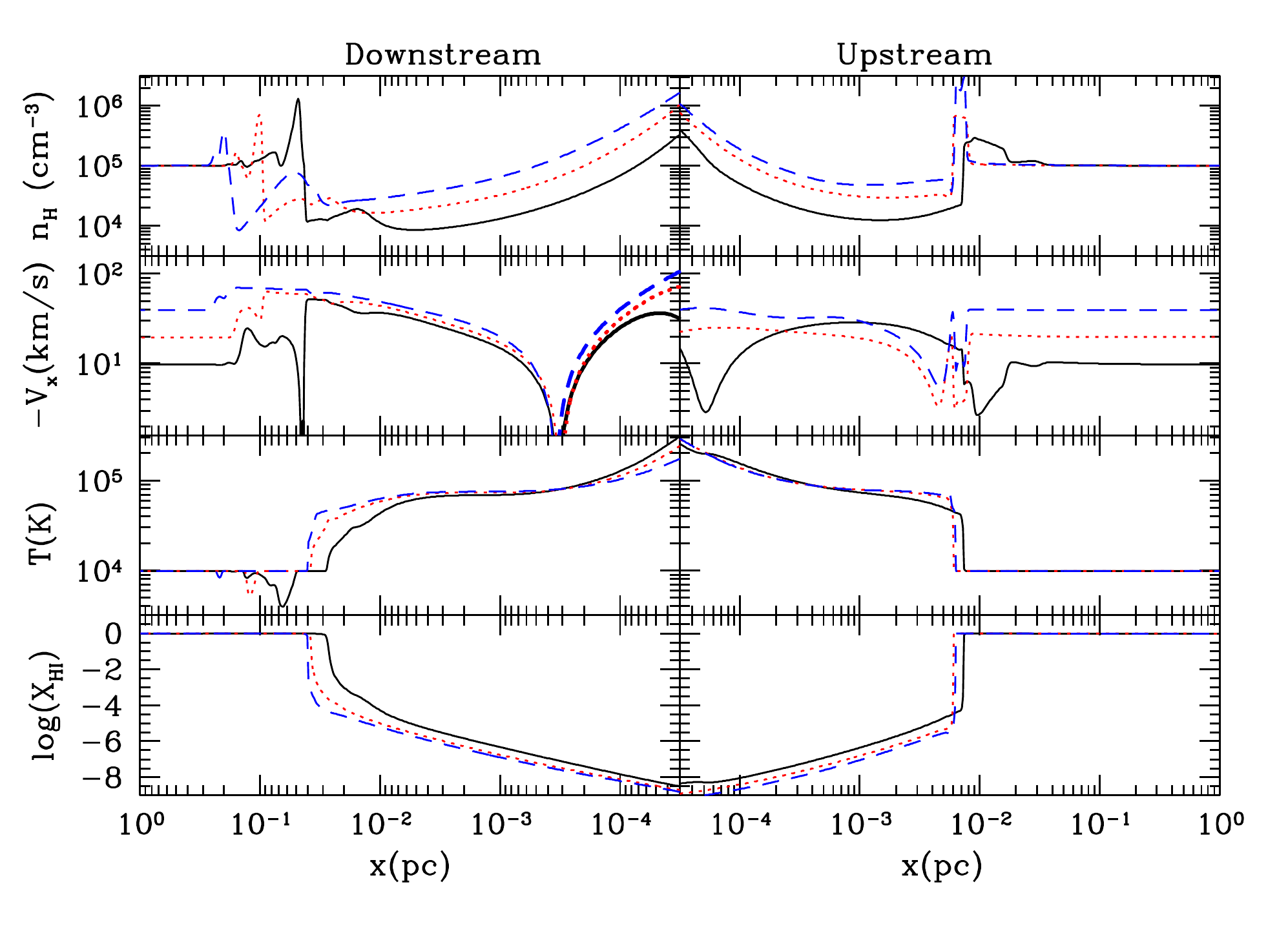}
\caption{Density, velocity, temperature, and \hi~ abundance (from top
to bottom) profiles along $X$-axis (axis of symmetry) for the simulations shown in Figure~\ref{den_d5}
for $\mach=1$ (solid lines), $2$ (dotted), and $4$ (dashed). The left panels
(note that small scales are on the right side) show profiles in
the downstream direction while the right panels show upstream profiles. }
\label{profile}
\end{figure*}

Figure~\ref{acc_d3} shows the time evolution of accretion rate for
different Mach numbers $\mach = 1,2,4,8,~$ and~$10$ for gas density
$\nH=\nthree$.  Most simulations, except the one with $\mach=4$, show
steady accretion rates after an initial transient phase that is the
result of out-of-equilibrium initial conditions.  The time duration of
the transient phase is proportional to the crossing timescale
$\tau_{\rm cr}$, and thus inversely proportional to the Mach number.
The mean dimensionless accretion rate $\lambdaavg$ (in units of the
Bondi rate) is measured when the accretion reaches a steady state
while for the non-steady cases, such as $\mach = 4$, we calculate the
time-averaged accretion rate. As found in Paper~I, at low gas
densities, i.e., $n_{\rm H, \infty} \lesssim \nfive$ for simulations
with $M_{\rm bh}=100~M_\odot$, $\lambdaavg$ is proportional to
square root of density ($\lambdaavg \propto n_{\rm H, \infty}^{1/2}$).
Figure~\ref{lambda} shows that after correcting $\lambdaavg$ for the
aforementioned density dependence also found for stationary BH, the
same functional form for the accretion rate fits all simulations with
different gas densities (large symbols for $\eta=0.1$ and gas
densities $\nH=10^2$--$\nfive$) and radiative efficiencies (small
pentagons for a simulation with $\eta=0.01$ and $\nH=\nfive$). Open
symbols indicate simulations that show non-steady accretion rate due
to instabilities of the dense post-shock layer forming in the upstream
direction (see Section~\ref{ssec:osc}).

The quasi-periodic oscillations of the accretion rate found in Paper~I
and Paper~II for stationary BHs are still observed in simulations with
low Mach numbers $\mach \lesssim 0.5$, which maintain the main
characteristics of spherically symmetric accretion discussed in
Paper~I and Paper~II. This implies that introducing small systematic
subsonic velocity to spherically symmetric accretion does not
significantly alter the oscillatory behavior of the
accretion. However, the average accretion rate $\lambdaavg$ decreases
steeply as a function of Mach number in this Mach number range, and
$\lambdaavg$ at $\mach \sim 1$ is roughly one order of magnitude
smaller than for non-moving BHs (including radiation feedback), and
three orders of magnitude smaller compared to stationary BHs with no
radiative feedback, when all the other parameters are held
constant. This decrease of the accretion rate with increasing velocity
found for subsonic BH velocities is only {\it qualitatively} similar
to Bondi--Hoyle--Lyttleton accretion, but does not have the same scaling
with BH velocity.

The spherically symmetric accretion model fails for supersonic BH
motion ($\mach \gtrsim 1$). The shape of the \hii~region makes a
transition to a well-defined axis-symmetric geometry, elongated along
the direction of the gas flow in the downstream direction, while a
bow-shaped dense shell develops in front of the \hii~region in the
upstream direction, significantly affecting the velocity field of the
gas inflow. In most simulations, steady-state accretion is achieved for
supersonic BH motion, since gas is continuously supplied to the BH
without interruption. 

Interestingly, as the BH motion becomes supersonic and a bow-shock and
dense shell form, $\lambdaavg$ increases as a function of Mach number.
This is clearly at odds with the results expected from the classical
Bondi--Hoyle--Lyttleton model. A Mach number of $\mach \sim 1$ is
roughly the turning point where $\lambdaavg$ has a minimum, while
there is a maximum value of the Mach number $\mach = \mach_R$
($\mach_R \sim 4$ for $T_\infty=10^4$~K), at which $\lambdaavg$
reaches a maximum value, before starting to decrease with increasing BH
velocity. An instability of the dense shell that leads to bursts of
accretion rate is observed in some simulations in this Mach number
range. This result will be discussed in Section~\ref{ssec:osc}. At
higher Mach numbers ($\mach > \mach_R$), a steady-state solution is
achieved once again since the dense shell does not form due to the
high velocity of the gas inflow (as shown in Section~\ref{ssec:dtype},
the I-front transitions from {\it D}-type to {\it R}-type). For $\mach
> \mach_R$, $\lambdaavg$ decreases monotonically as a function of Mach
number and converges to the Bondi--Hoyle--Lyttleton solution (with no
radiative feedback) shown as a dashed line in Figure~\ref{lambda}. In
this high velocity regime, the I-front becomes {\it R}-type and the
gas flow is weakly affected by radiative feedback.

\begin{figure*}[th]
\epsscale{1.1} 
\plottwo{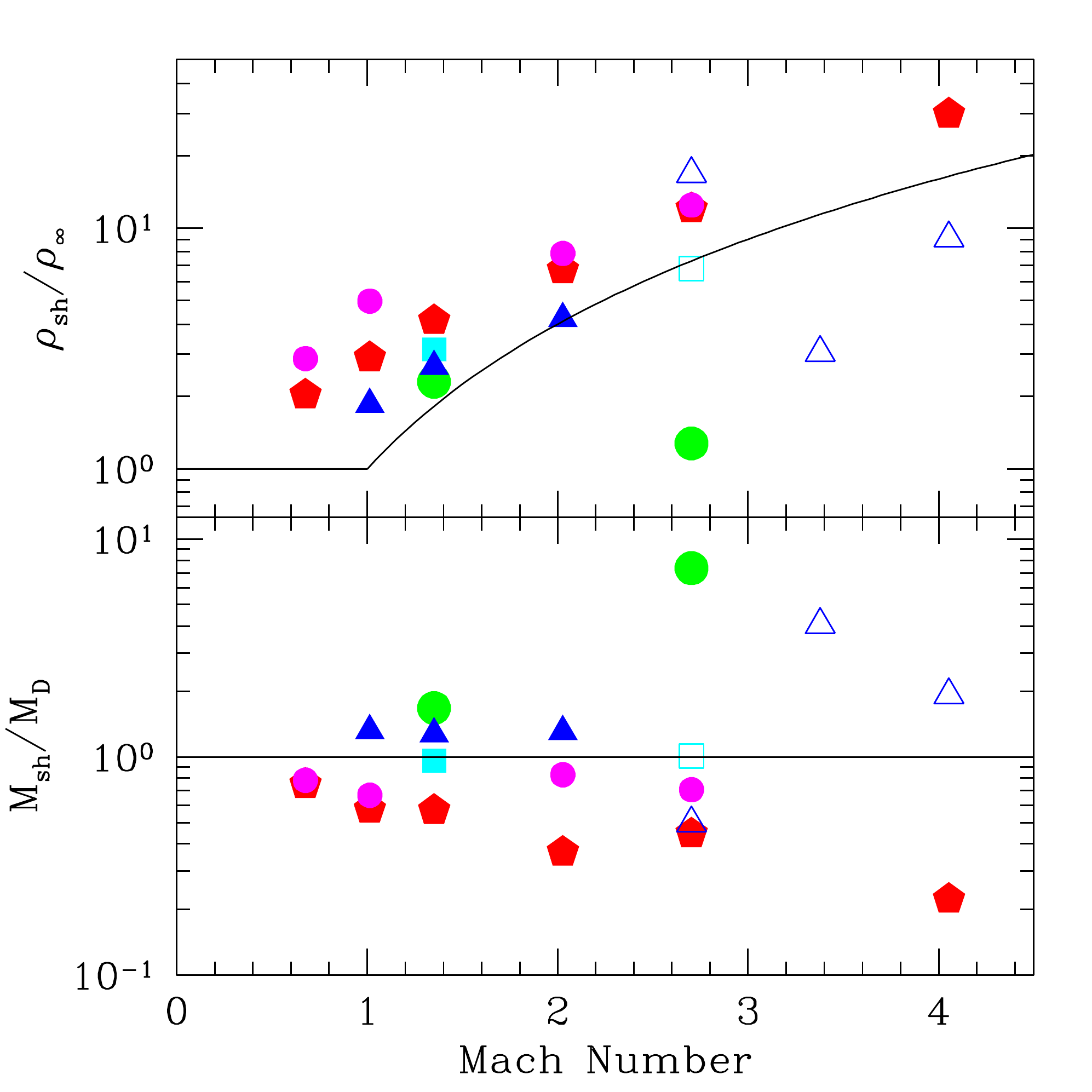}{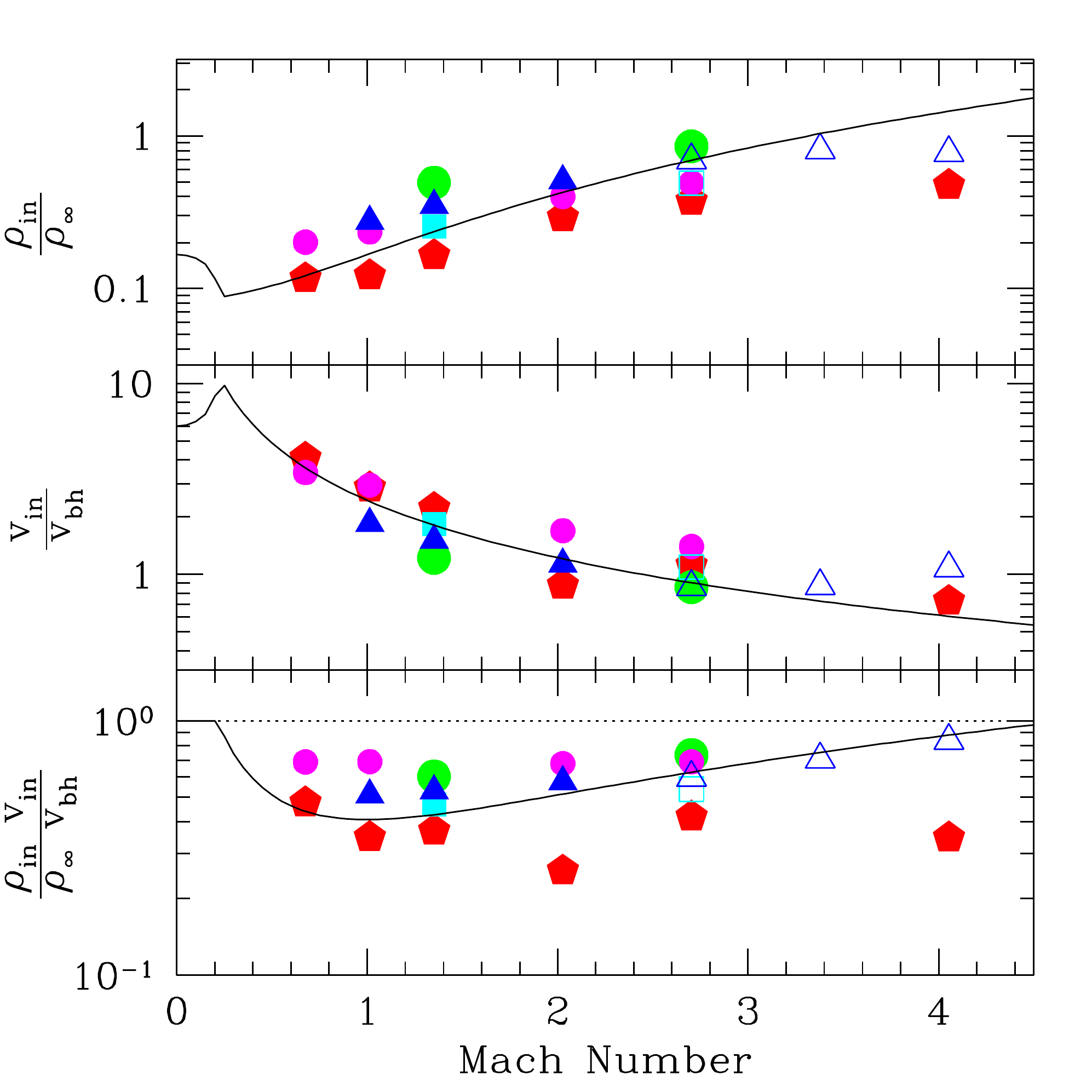}
\caption{{Left:} density (top panel) and Mach number (bottom
  panel) of gas in the dense shell in the upstream direction in the
  simulations shown in Figure~\ref{lambda}. The lines show the model
  predictions based on isothermal shock jump conditions ($\rho_{\rm
    sh}/\rho_\infty \propto \mach^{2}$) for the curved bow shock (see
  the text). {Right:} density, velocity and mass flux of gas inside
  the \hii~region in various simulations. The lines show the model
  predictions assuming {\it D}-type I-front (see the text). The symbols
  have the same meaning as in Figure~\ref{lambda}.}
\label{nv_in_out}
\end{figure*}

\subsection{Structure of the Gas Flow and {\hii}~Region}

For Mach numbers $1 < \mach <\mach_R$, a dense bow shock forms in
front of the \hii~region in the upstream direction, followed by a
{\it D}-type I-front (see Figure~\ref{den_d5}). Most of the gas inflow
propagates through the bow shock without changing direction, while a
small fraction of the gas inflow in the outer parts of the bow shock
is re-directed farther from the axis of symmetry. The formation of a
bow shock in the upstream direction changes the gas density and
velocity behind the shock, while the gas temperature remains
relatively unaffected (isothermal shock). Note that the \hii~region
has a cometary shape, with overall length increasing linearly with
increasing Mach number. The size of the \hii~region in the upstream
direction is not sensitive to the Mach number, while in the downstream
direction the length of the ionized tail shows a linear relationship
with the Mach number as shown in Figure~\ref{den_d5}.  The upper
panels in Figure~\ref{den_d5} show the changes in the density
structure and the \hii~region shape for $\mach = 1,2,~$and$~4$,
respectively. The lower panels show the vector fields over the gas
density for each simulation. In the bottom panels, we use a
logarithmic scale for the radial direction to better show the motion
of gas in the vicinity of the BH. The location Bondi radius calculated
for the gas temperature inside the \hii~region is close to, but within
the location of the I-front. We will see in the next section that this
will allow us to derive simple analytical formulae for the accretion
rate using the Bondi formula for the gas inside the \hii~region and a
model for the I-front.

For $\mach =1$, the size of the \hii~region in the downstream direction
is roughly $\sim 4$ times its length in the upstream direction. The
density structure in the downstream direction is very complex as shown
in Figure~\ref{den_d5}. The re-directed gas streams form high-density
regions and shocks. However, since most of the gas downstream of the
BH is not accreted onto the BH, we will focus on understanding the
upstream structures. The size of the \hii~region in the upstream direction 
will be discussed in greater detail in Section~\ref{ssec:hiisize}.

\section{Analytic Modeling}
\label{sec:modelling}

In this section we show that properties of the gas flow around the
moving BH can be understood using a simple one-dimensional model for a standing
I-front in the frame of reference moving with the BH. There are two
regimes for the I-fronts: an {\it R}-type front and {\it D}-type front, 
determined
by the BH velocity and density of the ambient medium. As in the
previous papers we find that the accretion rate onto the BH can be
estimated using the Bondi--Lyttleton formula for the gas inside the
I-front, since the effective Bondi radius lays within the ionized
region.

\subsection{Transition from {\it R}-type to {\it D}-type I-front}
\label{ssec:dtype}

In most simulations, we found that the gas flow reaches a steady-state
solution. In these cases, the position of the I-front with respect to
the BH is stationary, hence the I-front propagates with respect to the
gas upstream with constant velocity $\vbh$. Although the geometry of the
\hii~region and the bow shock are clearly not planar, we can
approximate the flow as one-dimensional (parallel to the direction of motion of the
BH) near the I-front location in the region around the axis of
symmetry of the problem. But we will show that accounting for deviations from the
planar symmetry is necessary to understand the simulations results.

Let us start by writing down the equations for the
propagation of one-dimensional I-front through an homogeneous medium
with density $\rho_\infty$. The density ratio
across the I-front can be estimated by solving the mass and momentum
conservation conditions $\rho_{\rm in} v_{\rm in} = \rho_\infty
\vbh = J\mu $, where $J$ is radiation flux and $\mu=1.27 m_H$,
assuming the helium becomes singly ionized in the front \citep{Spitzer:78}:
\begin{equation} 
\Delta_\rho \equiv \frac{\rho_{\rm in}}{\rho_{\rm \infty}} =
\frac{(1+\mach^2) \pm \sqrt{ (1+\mach^2)^2 -4 \mach^2 \Delta_T}} {2 \Delta_T}.  
\label{eq:rhoR}
\end{equation}
Here, we have defined $\Delta_T \equiv T_{\rm in}/T_{\rm \infty} \ge
1$. Due to the condition for the density ratio in
Equation~(\ref{eq:rhoR}) to have real positive values, the Mach number
must be $\mach\le \mach_D$, where {\it D} stands for {\it dense} gas,
or $\mach \ge \mach_R$ where {\it R} refers to {\it rarefied}
gas. {\it D}- and {\it R}-critical Mach numbers are respectively:
\begin{eqnarray}
\mach_D &=& \sqrt{\Delta_T}~(1-\sqrt{1-1/\Delta_T})~_{\overrightarrow{~\Delta_T \gg 1~}}~ \frac{1}{2 \sqrt{\Delta_T}},  \\
\mach_R &=& \sqrt{\Delta_T}~(1+\sqrt{1-1/\Delta_T})~_{\overrightarrow{~\Delta_T \gg 1~}}~2\sqrt{\Delta_T}.
\label{eq:machsols}
\end{eqnarray}
Also, $\mach_D \mach_R=1$, thus $\mach_D \equiv \mach_R^{-1}$. In
terms of the BH velocity the critical velocities are:
\begin{eqnarray}
v_D &=& c_{\rm s, in}~(1-\sqrt{1-1/\Delta_T})~_{\overrightarrow{~\Delta_T \gg 1~}}~ \frac{c_{\rm s,\infty}^2}{2 c_{\rm s, in}}, \\
v_R &=& c_{\rm s, in}~(1+\sqrt{1-1/\Delta_T})~_{\overrightarrow{~\Delta_T \gg 1~}}~2c_{\rm s,in}.
\label{eq:velsols}
\end{eqnarray}
Only for the special case in which $\Delta_T \simeq 1$, we have
$\mach_D \simeq \mach_R \simeq 1$ and a solution exists for any BH
velocity. However, for all physically motivated cases with $\Delta_T > 1$
a solution is not possible for BH velocities $v_D < \vbh < v_R$. In
our simulations if $\vbh >v_R \sim 2 c_{\rm s,in}$ the analytical
solution that reproduces the data is the one with the negative sign in
Equation~(\ref{eq:rhoR}). This solution describes a weak {\it R}-type I-front,
that has $\rho_{\rm in} \sim \rho_{\infty}$ and supersonic motions of the
gas both ahead and behind the I-front (in the frame of reference comoving
with the BH).

If $v_D < c_{\rm s,\infty}/2 < \vbh < v_R \sim 2 c_{\rm s,in}$ a solution
is not possible.  A shock front must precede the I-front, increasing
the density and reducing the gas velocity below $v_D$ ({\it D}-type
solution). This is indeed observed in the simulations. For a {\it D}-type
front, the density past the front is always lower than the density
upstream. Figure~\ref{profile} shows that the post-shock density and
velocity are a function of the Mach number, while the temperature of
the shell is always $T_{\rm sh} \approx T_\infty$, i.e., the shock is
isothermal ($\gamma=1$). The ratio between the densities at
infinity and behind the isothermal shock is
\begin{equation} 
\frac{\rho_{\rm sh}}{\rho_{\infty}} \simeq \mathcal{M}^2, 
\label{eq:rhoI}
\end{equation}
that is in good agreement with the simulation results (see
Figure~\ref{nv_in_out}, left panel).  However, for the calculation of
the gas velocity in the shell our simplifying assumption of planar
geometry fails. Assuming planar geometry, mass flux conservation in
the direction of the BH motion implies $\rho_{\rm sh} v_{\rm sh} =
\rho_\infty \vbh $. Thus, the Mach number in the shell $\mach_{\rm
  sh} \equiv v_{\rm sh}/c_{\rm s, sh}$ is $\mach_{\rm sh} =
\mathcal{M}^{-1} < 1$.  However, this assumption fails to reproduce
the simulation results and is inconsistent with obtaining a {\it D}-type
solution for the I-front. Hence, the model must assume (as verified in
the simulations) that the gas velocity inside the shell has a
non-zero tangential component. It follows that the component of the
velocity along the direction of motion of the BH is reduced due to
off-axis motions that divert some of the gas in the direction parallel
to the bow shock.
As illustrated in left panel in Figure~\ref{nv_in_out}, the simulations 
do not indicate a strong trend of $\mach_{\rm sh}$ with $\mach$:
\begin{equation}
\mach_{\rm sh} \sim {\cal A}\mach_D,~{\rm with}~{\cal A} \lesssim 1, 
\label{eq:vsh}
\end{equation}
implying that $v_{\rm sh}\rho_{\rm sh}/(\rho_\infty \vbh)={\cal A}(\mach/\mach_R)<1$.
Note that in order to be consistent with our model, the data points
for $\mach_{\rm sh}$ are calculated from the simulation data by
enforcing mass conservation across the I-front: $\mach_{\rm
  sh}=\rho_{\rm in} v_{\rm in}/(\rho_{\rm sh}c_{\rm s, sh})$. This
ensures that the jump conditions across the I-front ignore the
component of the velocity parallel to the bow shock. 

\begin{figure}[t] 
\epsscale{\myscale} 
\plotone{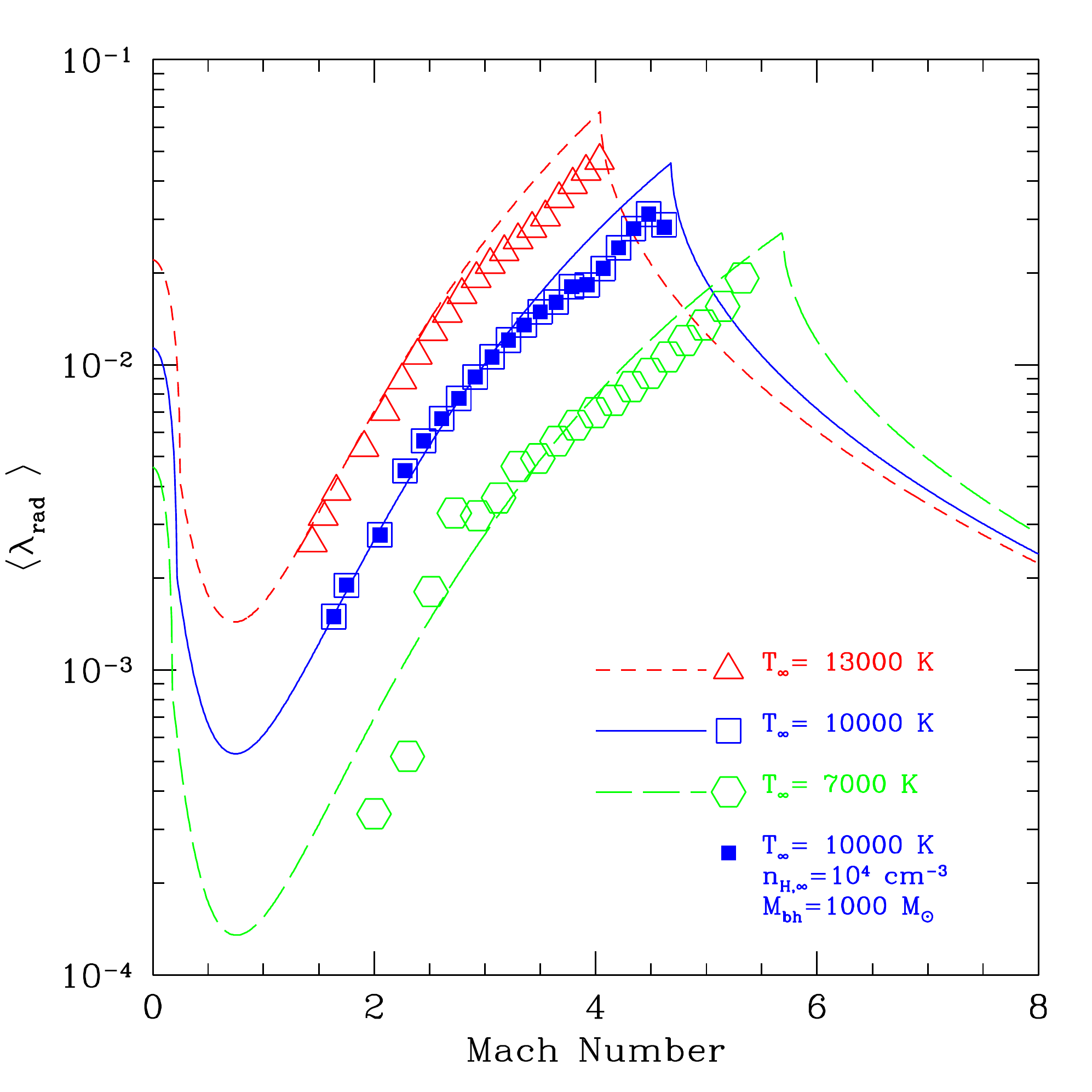}
\caption{Accretion rates as a function of Mach number for
  ``wind-tunnel'' simulations with $M_{\rm bh}=100~M_\odot$,
  $\nH=\nfive$, and $\eta=0.1$. Different lines and symbols show the
  model and simulations respectively for gas temperatures at infinity
  $T_\infty=7,000~$K (long-dashed line, open hexagons), $10,000$~K
  (solid line, open squares), and $13,000$~K (short-dashed line,
  triangles). Solid squares show simulations with $T_\infty=10,000$~K
  but BH mass 10 times higher ($M_{\rm bh}=1000~M_\odot$) and ambient
  gas density 10 times lower ($\nH=\nfour$) than the other runs in
  this figure. The accretion rates for the two simulations with
  $T_\infty=10,000$K (solid and open squares) are indistinguishable.}
\label{accel}
\end{figure}

As discussed above, across the I-front, the density ratio between the
gas in the shell and in the $\hii$~region can be estimated by solving
the mass and momentum conservation conditions:
\begin{eqnarray} 
\frac{\rho_{\rm in}}{\rho_{\rm sh}} &=&
\frac{(1+\mach_{\rm sh}^2) \pm \sqrt{ (1+\mach_{\rm sh}^2)^2 -4 \mach_
{\rm sh}^2 \Delta_T}} {2 \Delta_T} \nonumber \\ &\sim& 2\mach_D^2~{\rm for}~{\cal A} \sim 1. 
\label{eq:rhoII}
\end{eqnarray}
For Equation~(\ref{eq:rhoII}) to have real positive values, we must
have $\mach_{\rm sh} \le \mach_D= \mach_R^{-1}$ ({\it D}-type
solution). Clearly, if ${\cal A}\le 1$ (i.e.\, $\mach_{\rm sh} \le
\mach_D$) a {\it D}-type solution is possible for any Mach number
$\mach<\mach_R$. Note that a {\it D}-type solution would not be
possible if we had assumed one-dimensional flow geometry for which $\mach_{\rm
  sh}=\mach^{-1}$. One-dimensional flow assumption would give a {\it D}-type
solution only for $\mach > \mach_R$, that is the regime in which we
expect an {\it R}-type front, and is thus ruled out. In general, for
${\cal A}^2 < 1$ the {\it D}-type solution consistent with the
simulations results is the one with the negative sign in
Equation~(\ref{eq:rhoII}), which has the larger relative decrease in
density across the front: $\rho_{\rm in}/\rho_{\rm sh} \sim {\cal
  A}^2\mach_D^2$. This solution describes a strong {\it D}-type front.
By combining Equations~(\ref{eq:rhoI})--(\ref{eq:rhoII}), we obtain 
\begin{eqnarray} 
\Delta_\rho &\equiv& \frac{\rho_{\rm in}}{\rho_{\infty}} =
\frac{\rho_{\rm in}}{\rho_{\rm sh}} \frac{\rho_{\rm sh}}{\rho_{\infty}} \\
&\approx & \frac{\mach^2} {2 \Delta_T} \approx 2\left(\frac{\mach}{\mach_R}\right)^2~{\rm for}~{\cal A} \sim 1, 
\label{eq:delta_rho}
\end{eqnarray}
or $\Delta_\rho={\cal A}^2 (\mach/\mach_R)^2$ for ${\cal A}^2<1$.
The velocity ratio between the gas inside the \hii~region and the
BH velocity is $ v_{\rm in}/\vbh = (\rho_{\rm sh}/\rho_{\rm
in})(\mach_{\rm sh}/\mach) \approx \mach_R/2\mach $ for ${\cal A}\sim 1$
or $v_{\rm in}/\vbh = \mach_R/{\cal A}\mach$ for ${\cal A}^2<1$. In this regime
({\it D}-type I-front) the model predicts that the Mach number of the gas
inside the \hii~region is a constant close to unity, as is indeed observed
in all the simulations: $\mach_{\rm in} \approx (v_{\rm in}/\vbh)
(\mach /\Delta_T^{1/2}) \simeq 1$ for ${\cal A}\sim 1$ (or $\mach_{\rm in} =
2/{\cal A}$ for ${\cal A}^2<1$).

\begin{figure}[t] 
\epsscale{\myscale} 
\plotone{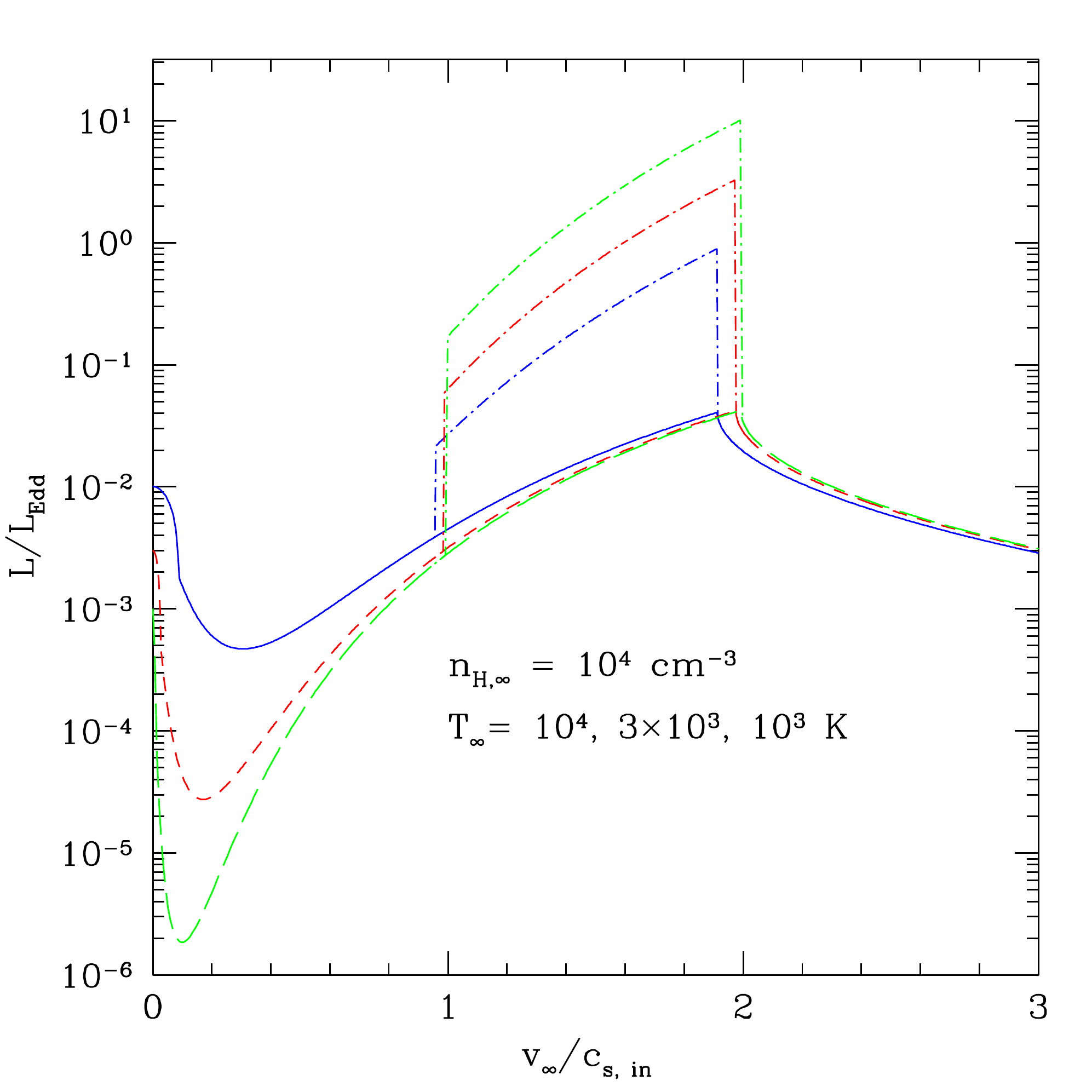} 
\caption{Accretion luminosity normalized by the Eddington rate as a
  function of the BH velocity $\vbh$ in units of the sound speed
  inside the \hii~region for BH mass $M_{\rm bh}=100~M_\odot$,
  $\nH=\nfour$, and $\eta=0.1$. Different line types are used for
  models with different temperatures: $T_\infty = 10^4$~{\rm
    (solid)},$~3\times 10^3$~{\rm (short dashed)},~{\rm and}~$10^3~$K~
  (long dashed). Note that the accretion rates peak at $\vbh \sim 2
  c_{\rm s,in}$ regardless of $T_\infty$. Dot-dashed lines for each
  temperature $T_\infty$ show possible enhanced luminosities when the
  thin shell in front of the I-front becomes unstable producing
  periodic accretion bursts. Accretion of high-density gas from the
  broken dense shell remnants with density $\rho_{\rm sh} \propto
  \mach^2$ can increase the peak accretion rates by a factor
  $\mach^2$. }
\label{low_temp}
\end{figure}
 
\begin{figure*}[t]
\epsscale{\myscale} 
\plotone{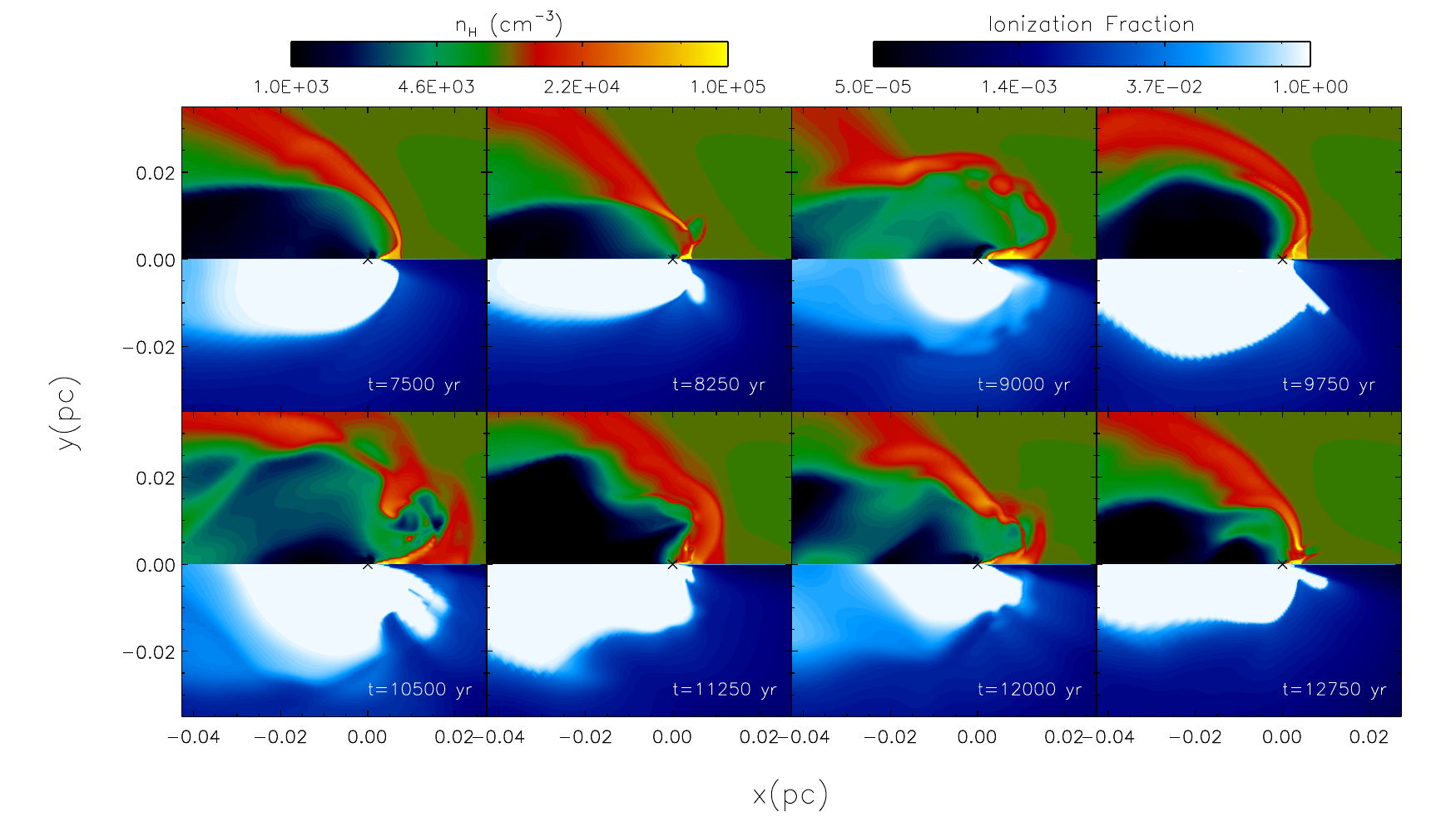}
\caption{Snapshots showing the time evolution of the density (top half
  of the panels) and ionization fraction (bottom half of the panels)
  in a simulation of a BH of mass $M_{\rm bh} = 100~M_\odot$, gas
  density $\nfour$, and temperature $T_{\infty}=10^4~$K moving at
  $\mach=2.7$. The dense gas shell upstream of the bow shock forms and
  gets destroyed with a quasi-periodic behavior. When the dense shell
  breaks and falls onto the BH, the accretion rate shows peak
  luminosities which are much higher than the average depending on the
  density of the shell, and thus the Mach number $\mach$.}
\label{fig:den_d4}
\end{figure*}

The symbols in the upper left panel in Figure~\ref{nv_in_out} show the
ratio between the gas density at infinity and inside the dense shell as
a function of $\mach$ for a set of simulations with different
parameters (i.e., gas density and radiative efficiency as shown in the
figure's legend). Our model is shown as a solid line. The lower left
panel shows the Mach number in the shell $\mach_{\rm sh}$ in units of
$\mach_D$, as a function of $\mach$ for the same simulations.  The
right panels in Figure~\ref{nv_in_out} show the density and velocity
ratios between the gas at infinity and inside the \hii~region as a
function of Mach number.  We measure the gas density and velocity
within the \hii~region where the density profiles have a minimum behind
the I-front.  As discussed above, for a plane parallel I-front and
shock the mass flux of the gas is conserved: $\rho_{\rm in} v_{\rm
  in}/(\rho_{\infty} \vbh)=1$. However, due to the formation of
a bow shock when the I-front becomes {\it D}-type, the gas flow has a
velocity component perpendicular to the direction of the BH motion and
$\rho_{\rm in} v_{\rm in}/(\rho_{\infty} \vbh) \approx {\cal A}
\mach/\mach_R< 1$, in agreement with the simulation results shown in
the right bottom panel of Figure~\ref{nv_in_out}.  In our fiducial
simulations with $T_\infty=10^4$~K, we have $\mach_{R} \sim 4.7$
($\Delta_T \simeq 6$ and $T_{\rm in}= 6 \times 10^4~{\rm K}$). The
simulation results confirm that a {\it D}-type I-front forms for
$\mach < \mach_R$, while a transition to {\it R}-type occurs at $\mach
\sim \mach_R$.

\subsection{Accretion Rate}

In this section, we describe a model that fits the gas accretion rates
measured at the inner boundary (i.e., near the BH) in all our
simulations. Since the effective Bondi radius calculated using the
velocity and the sound speed inside the \hii~region is smaller than
the radius of the I-front in the upstream direction, we can treat this
problem as a ``mini" Bondi accretion within the \hii~region.  The mean
accretion rate can be estimated using the Bondi--Hoyle--Lyttleton
formula for gas inside the \hii~region: $\dot{M} \propto M_{\rm bh}^2
\rho_{\rm in} c_{\rm s, in}^{-3} (1+\mathcal{M}_{\rm
  in}^2)^{-3/2}$. As in previous papers of this series, we define the
dimensionless accretion rate in units of the Bondi rate $\dot{M}_B
\propto M_{\rm bh}^2 \rho_{\infty} c_{\rm s, \infty}^{-3}$:
\begin{eqnarray}
\lambdaavg \equiv \frac{\dot{M}}{\dot{M}_B} &=& \frac{\rho_{\rm
in}}{\rho_{\infty}} \left( \frac{c_{\rm s, \infty}}{c_{\rm s, in}}
\right)^3 \frac{1}{(1+\mathcal{M}_{\rm in}^2)^{3/2}} \\  
&=& \frac{\Delta_\rho}{\Delta_T^{3/2}} \frac{1}{(1+\mathcal{M}_{\rm in}^2)^{3/2}}. 
\end{eqnarray}
We have identified two regimes corresponding to the {\it D}-type and
{\it R}-type I-front solutions.  If $1 < \mach \le \mach_R$, in the
previous section we found $\Delta_\rho \approx 2(\mach/\mach_R)^2 =
2(\vbh/v_R)^2$, and $\mach_{\rm in} \approx 1$. This means that inside
the \hii~region the density increases as the BH moves faster but the
Mach number remains constant (approximately transonic). This model
explains in simple terms the increase in the accretion rate with
increasing BH velocity observed in the simulations in this regime. If
$\mach > \mach_R$, the shock front does not form and $\Delta_\rho
\rightarrow 1$ (if $\mach \gg \mach_R$), while the Mach number is
$\mach_{\rm in} \approx \mach/\Delta_\rho/\Delta_T^{1/2}$.  Hence, in
dimensionless units:

\begin{equation}
\lambdaavg \approx  
\begin{cases}
(2\Delta_T)^{-5/2}\mach^2 \approx \frac{2^{5/2}}{\mach_R^3}(\frac{\mach}{\mach_R})^2, \\ \hfill \text{if~$\mach \le \mach_R$}~~~~~~~~~~ \\
\Delta_\rho^{4}(\Delta_\rho^2\Delta_T+\mach^2)^{-3/2}_{\overrightarrow{\mach\gg\mach_R}} [(\mach_R/2)^2+\mach^2]^{-3/2}, \\ \hfill \text{if~$\mach > \mach_R$}.~~~~~~~~~~
\end{cases}
\end{equation}

The solid line in Figure~\ref{lambda} shows our model for the
accretion rate onto moving BHs. The model generally is a good fit to
the simulation results, except the model slightly overestimates the
accretion rate around the critical Mach number $\mach_R$.

We run a complementary set of simulations to study more precisely the
changes of the physical properties as a function of Mach number, since
the simulations with constant velocities have a coarse sampling in
velocity space and show an intrinsic scatter which is probably the
result of out-of-equilibrium initial conditions.  We start the
simulation assuming sonic motion of the BH ($\mach \sim 1$) and
increase gradually the velocity of the gas inflow at the
boundary. This type of ``{\it wind-tunnel}" numerical experiments is
useful to focus on the changes of physical properties as a function of
velocity, while holding the other parameters fixed. The critical Mach
number $\mach_R$ and the peak luminosity depend on the temperature
ratio $\Delta_T$ as in
Equation~(\ref{eq:machsols}). Figure~\ref{accel} shows the accretion
rate as a function of Mach number for different gas temperatures at
infinity $T_\infty = 7\times10^3, 10^4$, and $1.3\times 10^4$~K for
$M_{\rm bh}=100~M_\odot$, $\nH=\nfive$, and $\eta=0.1$. Our model is
in good agreement with the simulation results as shown in
Figure~\ref{accel}. The peak accretion rates and the critical Mach
numbers in these cases are very close to the model
predictions. However, the caveat is that the dense shell which
initially forms at the beginning of the simulations does not change
its location as the velocity of gas increases as observed in the
simulations in which $\vbh$ was held constant.

If we express the accretion rate as a function of the BH velocity in
physical units, we find that the accretion rate is independent of the
temperature of the ambient medium and peaks at about twice the sound
speed inside the \hii~region: $\vbh^{\rm max}=v_R \simeq 2 c_{\rm s,
  in}$ (see Figure~\ref{low_temp}). This is contrary to the case of a
static BH (or moving at subsonic speed) for which $\lambdaavg \propto
T_\infty^{5/2}$ and hence the accretion rate ${\dot M} \propto
T_\infty$ is proportional to the temperature of the ambient gas. The
gas accretion rate for a moving BH is
\begin{equation}
{\dot M} \approx \frac{\rho_\infty (G M_{\rm bh})^2}{c_{\rm s,in}^3} \times
\begin{cases}
0.7\left(\frac{\vbh}{2c_{\rm s,in}}\right)^2~\text{if~$c_{\rm s,\infty}<\vbh \le 2c_{\rm s,in}$}\\
[1+(\vbh/c_{\rm s,in})^2]^{-3/2}~\text{if~$\vbh \gg 2 c_{\rm s, in}$}.
\end{cases}
\label{eq:res}
\end{equation}
The velocity for peak accretion depends only on the sound speed inside
the \hii~region and is $v_R=50~{\rm km~s^{-1}}$ for $T_{\rm
  in}=6\times10^4~{\rm K}$ ($c_{\rm s,in}=25~{\rm km~s^{-1}}$).
We find a mild dependence of $T_{\rm in}$ on the density of the
ambient medium.  As explored in more detail in Paper~I, $T_{\rm in}$
depends on the hardness of the spectrum emitted by the BHs, the gas
metallicity, and Compton cooling/heating.

Our model has a peak accretion rate of about $70\%$ of the Bondi rate
in a gas with temperature $T_{\rm in}$ and density $\rho_\infty$, in
good agreement with our ``{\it wind- tunnel}'' type simulations
(however, as noted before, the simulations with constant BH Mach
number produce smaller accretion rates at peak accretion: about 
a factor of five smaller than the model prediction for the peak value).
Interestingly, in our model and simulations a BH
moving at $20$--$50~{\rm km~s^{-1}}$ with respect to a gas with temperature
$T_\infty \lesssim 10^4$~K, has a faster growth rate and accretion
luminosity than if it was at rest (or moving at subsonic speed).
Figure~\ref{low_temp} is an extrapolation of our results to lower
temperature regime showing accretion rates as a function of BH
velocity $\vbh$ for gas temperatures $T_\infty = 10^4, 3\times
10^3, {\rm and}~10^3$~K. In all cases, if $T_\infty < 10^4~$~K, we
find that the peak accretion rate at $\vbh \sim 2c_{\rm s, in}$
is larger than the corresponding accretion rate onto a stationary BH.
Clearly this is an important result because significant BH accretion
is only possible when the BH is in a dense medium, for instance, a
molecular cloud or the cold neutral medium in a galaxy, that
generally have low temperatures. In addition, the colder the gas the
smaller the BH velocity needs to be to achieve the supersonic speed
that leads to significant increase of the BH accretion rate.

In the next section, we will discuss another important result: in the
regime when the I-front is {\it D}-type, instabilities of the dense
and thin shell behind the bow shock may lead to its fragmentation and
hence produce periodic oscillations of the accretion rate and
luminosity of the BH. This effect can further increase the peak
accretion luminosity of the BH by roughly a factor $\mach^2$ with
respect to the mean values estimated above in Equation~(\ref{eq:res}).
For instance, the simulation with $\mach \sim 3$, $\nH=\nfour$, and
$T_\infty = 10^4$~K shows this instability and the peak accretion
rate is roughly a factor of 10 larger than the mean.

\subsection{Stability of Bow Shock and Periodic Oscillations of the Luminosity} 
\label{ssec:osc}
As discussed in the previous section, the average accretion rate
$\lambdaavg$ increases with increasing Mach number if
$1<\mach<\mach_R$. For the lower values in this Mach number range,
all simulations approach a steady-state accretion rate. Interestingly,
as the Mach number approaches $\mach_R$, the bow shock in simulations
with ambient gas densities in the range $\nH=10^3$--$\nfour$, becomes
unstable producing intermittent bursts of accretion due to a cyclic
formation/destruction of a dense shell in the upstream direction
\citep*[see][]{WhalenN:11}.  Figure~\ref{fig:den_d4} shows time
evolution of a simulation with $M_{\rm bh} = 100~M_\odot$, gas density
$\nH=\nfour$, temperature $T_{\infty}=10^4~$K, and Mach number
$\mach=2.7$.  As seen in Figure~\ref{den_d5}, the ionizing radiation
creates a ``cometary-shaped'' \hii~region around the BH. In the early
stages of the simulation, the gas flow remains relatively
steady. However, with time some instabilities start growing leading to
the fragmentation of the shell.  As a result, the ionizing radiation
and the hot gas inside the \hii~region are no longer contained by the
dense shell downstream of the bow shock and an explosion takes
place. Fragments from the broken dense shell fall onto the BH
significantly increasing the accretion rate, thus creating more
ionizing photons that blow out further the thinner parts of the
shell. However, after a time delay from the burst roughly estimated as
the \hii~region sound crossing time, the dense shell re-forms,
resetting the initial conditions for the next burst cycle.

Since the dense shell is located outside the inner Bondi radius, the
gravitational acceleration on the shell is relatively small and hence
the time scale for Rayleigh--Taylor instability is long. In addition
the radiation has a stabilizing effect as it tends to smooth out the
growth of linear perturbations on small scales, that have the faster
grow rate. However, when $\vbh \gtrsim c_{\rm s, in}$ (i.e., $\mach
\sim \mach_R/2$) the ram pressure becomes comparable to the thermal
pressure inside the \hii~region and the I-front is pushed closer to
the BH. In this case the increased gravitational acceleration on the
shell and the increased sharpness of the pressure gradient seem to
trigger the growth of instabilities. Also the column density of dense
shell is roughly constant as a function of the BH velocity, hence
because $\rho_{\rm sh} \propto \mach^2$ the thickness of the dense
shell is $\propto \mach^{-2}$. A thin shell is well known to be
unstable \citep{Vishniac:94,WhalenN:08a, WhalenN:08b}.
 
\begin{figure}[t]
\epsscale{\myscale} 
\plotone{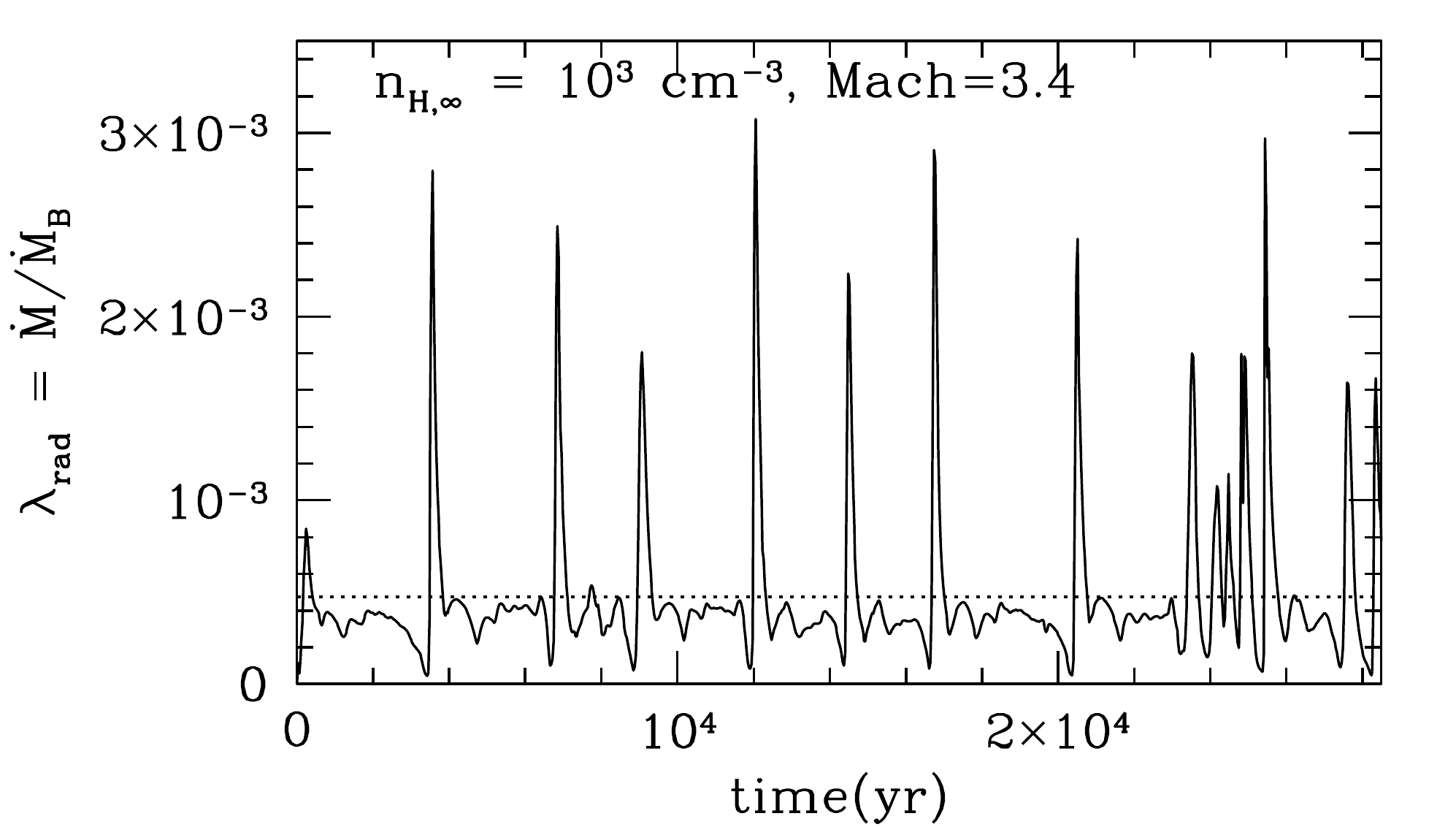}
\caption{Accretion rate in units of the Bondi rate as a function of
  time in a simulation of a BH with mass $M_{\rm bh} = 100~M_\odot$,
  gas density $\nthree$, and temperature $T_{\infty}=10^4~{\rm K}$
  moving at $\mach=3.4$. In this simulation, the dense gas shell
  upstream of the bow shock is unstable but after its fragmentation it
  re-forms on a timescale comparable to the \hii~region sound
  crossing time.
When the dense shell breaks and falls onto the BH, the accretion rate
shows peak luminosities which are an order of magnitude higher than
the average (shown as dotted line).}
\label{acc_d3m500}
\end{figure}

As shown in Figure~\ref{nv_in_out}, in simulations at intermediate
densities ($\nH=10^3$--$\nfour$) the Mach number in the dense shell is
close but slightly larger than $\mach_D$ while for higher densities
($\nfive$) is always less than $\mach_D$. This implies that {\it
  D}-type solution for $\nfive$ exists for all $\mach$ since
$\mach_{\rm sh} < \mach_D$ while a transition between {\it D}- and
{\it R}-type is expected for $\nH=10^3$--$\nfour$.  As shown in
Figure~\ref{acc_d3m500}, for simulations with $\nH=\nthree$ the
interval between the bursts of accretion is fairly constant $\Delta t
\sim 3000$ years.

In Paper~I and Paper~II we found a linear relationship between the
average size of the \str spheres $\ars$ and the period between
oscillations for stationary BHs, which is of a few $10^3$ years for
$M_{\rm bh} = 100~M_\odot$ and $\eta=0.1$.  We have shown that this
time scale is proportional to the sound crossing time of the
\hii~region and have provided analytical relationships for the
period. The same argument can be applied for interpreting the period
between intermittent bursts for moving BHs and in the next section we
provide a model that describes the size and shape of the \hii~region
around moving BHs.

\subsection{Size of {\hii}~Region in the Up/Downstream} 

\begin{figure}[t]
\epsscale{\myscale} 
\plotone{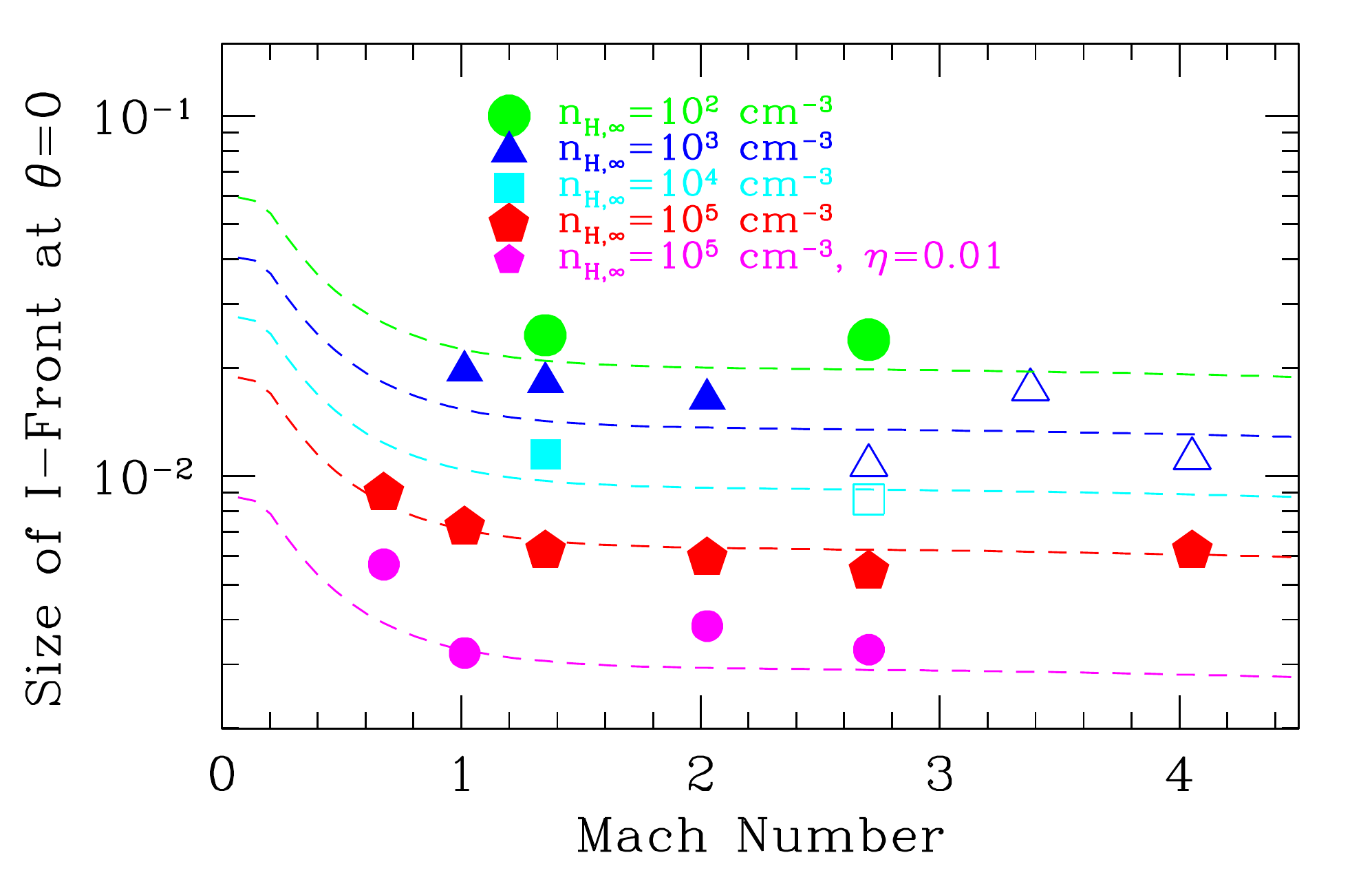}
\caption{Size of \hii~region in the upstream direction ($\theta=0$)
  $\ars_{\theta=0}$ as a function of $\mach$.  Simulation results
  (symbols) show a good match with the model (lines). The symbols have
  the same meaning as in Figure~\ref{lambda}. Since the recombination
  timescale $\tau_{\rm rec}$ within the \hii~region is shorter than
  the \hii~region crossing time, $\ars_{\theta=0}$ is not sensitive to
  the velocity of the flow. Also, $\ars_{\theta=0}$ is very not
  sensitive to the density of the ambient medium since
  $\ars_{\theta=0} \propto \nH ^{-1/6}$.}
\label{if0}
\end{figure}

The gas inflow in the direction of the polar axis $\theta=0$
(upstream) and $\theta=\pi$ (downstream) can be approximated by a one-dimensional 
flow with the \hii~region being supplied with neutral gas with
constant velocity $\vbh$. The total number of ionizing photons
emitted by the BH must equal the number of hydrogen recombinations
inside a radius $\ars$ in addition to the flow through the I-front of
neutral gas of density $n$ and velocity $\vbh$:
\begin{equation} 
N_{\rm ion} =
\frac{4\pi}{3} \ars_{\rm \theta}^3 \alpha_{\rm rec} n_e^2 + 4\pi 
\ars_{\rm \theta}^2 n \vbh \cos(\theta),
\label{eq:N0}
\end{equation}
where $N_{\rm ion}$ is the number of emitted ionizing photons, being
directly related to the luminosity of the BHs (which is a function of
Mach number). When the magnitudes of the two terms on the right side
of Equation~(\ref{eq:N0}) are compared, at $\theta=0$ the first
term is dominant over the second term due to the BH motion. The
electron number density inside the \hii~region for $1< \mach <
\mach_R$ is
\begin{equation} 
n_e \sim x_e n_{\rm H,in} = \frac{\mach^2}{(2\Delta_T)^{5/2}} \nH. 
\end{equation}
Since $\lambdaavg \propto \nH^{1/2}$ (for $\nH \lesssim
\nfive$ and $M_{\rm bh}=100~M_\odot$), the average
size of the \hii~region in the upstream direction is
\begin{equation} 
\ars_{\rm \theta=0} \propto \eta^{1/3} \nH^{-1/6}.   
\end{equation}
Figure~\ref{if0} shows the size of \hii~region at $\theta=0$ as a
function of Mach number for simulations with various densities ($\nH =
10^2$--$\nfive$) and radiative efficiencies ($\eta=0.1, 0.01$). The model
is a good fit to the simulation results.


\begin{figure}[t]
\epsscale{\myscale} 
\plotone{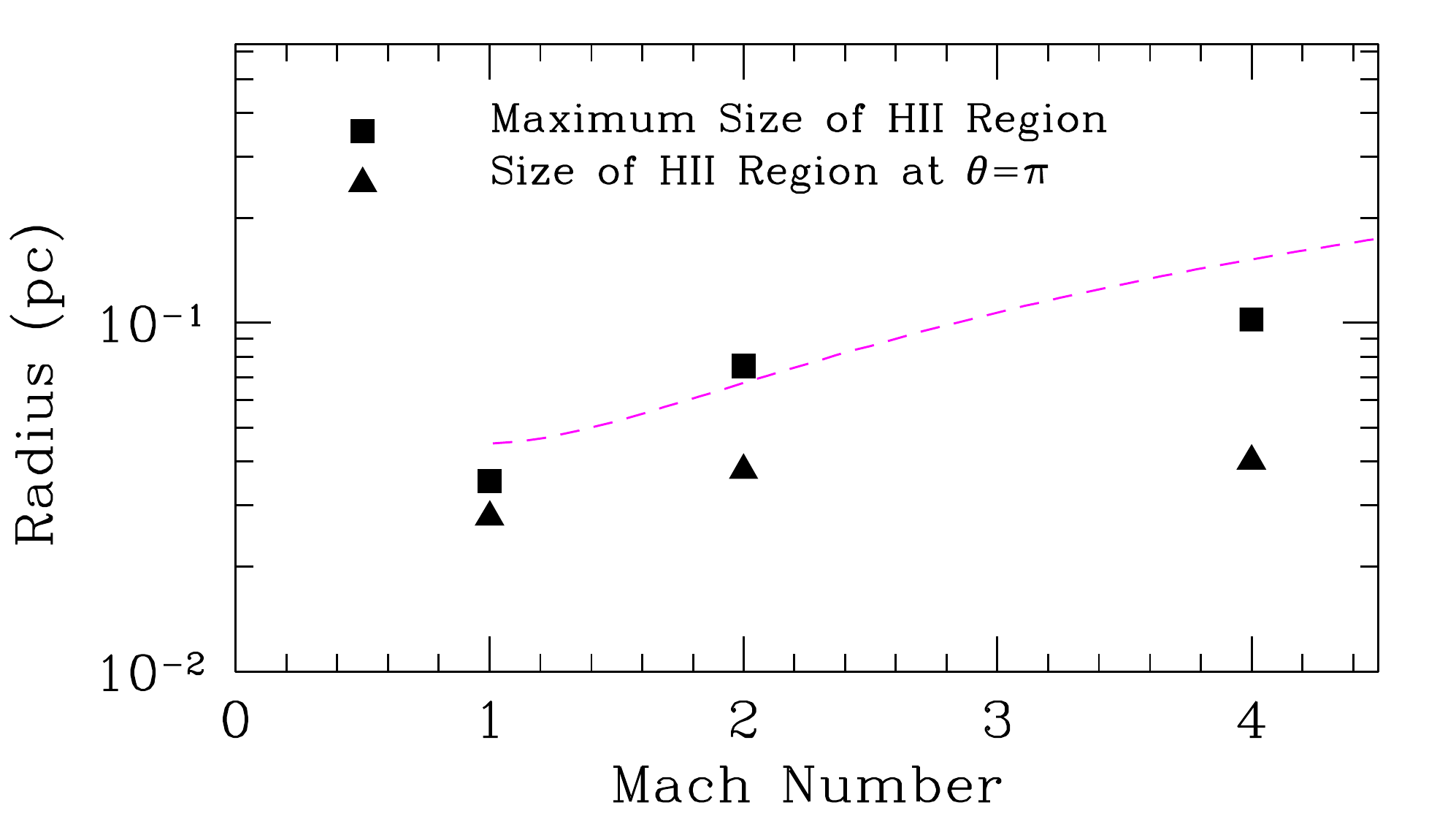}
\caption{Size of \hii~region in the downstream direction as a function
  of $\mach$. Maximum size of \hii~region and $\ars_{\rm \theta=\pi}$
  are shown as squares and triangles, respectively. Our model shown as
  a dashed line approximately reproduces the maximum size of the
  \hii~region in the downstream direction (that has an offset from the
  axis of symmetry). The size of the \hii~region in the direction
  $\theta=\pi$ is smaller than the maximum size of the \hii~region due
  to gravitational focusing of gas by the BH that enhances the gas
  density along the axis (at $\theta=\pi$).}
\label{if_pi}
\end{figure}

We model the size of the \hii~region in the downstream direction
in a similar manner. In the downstream direction $\theta=\pi$ the
second term on the right-hand side of Equation~(\ref{eq:N0}) dominates
over the first term:
\begin{equation} 
N_{\rm ion} \simeq 4\pi \ars_{\rm \theta=\pi}^2 n_{\rm H,in} \vbh, 
\label{eq:N0_down}
\end{equation}
where $n_{\rm H,in}$ can be calculated simply using pressure
equilibrium condition $n_{\rm H,in} = n_{\rm H,\infty}
\Delta_T^{-1}$. The size of the \hii~region in the downstream
direction for $1 < \mach < \mach_R$ is
\begin{eqnarray}
\ars_{\rm \theta=\pi} \propto \eta^{1/2} \nH^{1/4} (1+\mach^2)^{1/2}, 
\end{eqnarray}
where $\ars_{\rm \theta=\pi}$ is approximately proportional to the
Mach number. Figure~\ref{if_pi} shows that the model reproduces the
length of the tail of the \hii~region in the downstream direction in
simulations for $\nH=\nfive$. The length of the \hii~region at
$\theta=\pi$ is shorter than the model prediction because of the
gravitational focusing of the gas by the BH that increases the gas
density on the axis of symmetry behind the BH.

\section{Summary and Discussion}
\label{sec:discussion}
\label{ssec:hiisize}

In this third paper of a series on radiation-regulated accretion onto
BHs, we have focused on the effect of the BH motion relative to the
surrounding gas, i.e., the Hoyle--Lyttleton problem modified by
the effects that photo heating and radiation pressure from the
radiation emitted near the BH have on the accretion flow. As in
previous papers of this series, we have used radiation-hydrodynamic
simulations to explore a large parameter space of initial conditions
to inform us on how to formulate an analytical model that reproduces
the simulation results. The followings are our key findings.

\begin{itemize}
\item The quasi-periodic oscillation of the accretion rate observed in
  simulations of non-moving BHs is only observed for subsonic motions of
  the BH, while the accretion rate becomes steady at supersonic velocities
  (in most cases).

\item In the supersonic regime, we observe an axis-symmetric gas flow
  and a ``cometary-shaped'' \hii~region with tail length proportional to
  the BH velocity $\vbh$. For BH velocities $c_{\rm s,\infty} <\vbh< v_R
  \approx 50~{\rm km~s^{-1}}$ the ionization front becomes {\it D}-type:
  a bow shock and a dense shell develop in front of the \hii~region in
  the upstream direction. For $\vbh > v_R$ the bow shock disappears and
  the I-front becomes {\it R}-type.

\item For subsonic motion of the BH we find that the accretion rate
  onto the BH decreases with increasing BH velocity. However, contrary
  to naive expectations, the accretion rate increases with increasing BH
  velocity in the regime $c_{\rm s,\infty}< \vbh < 50~{\rm km~s^{-1}}$,
  when the I-front is {\it D}-type. The accretion rate peaks at
  BH velocity $\vbh \approx 50~{\rm km~s^{-1}}$ before it starts
  decreasing with increasing BH velocity, converging to the well-known
  Hoyle--Lyttleton solution without radiation feedback.

\item Based on the simulation results, we formulate a simple analytical
model of the problem based on modeling the jump conditions across the
I-front. The transition of the I-front from {\it R}-type to {\it D}-type
happens at $\vbh=2 c_{\rm s,in}$, where $c_{\rm s,in}$ is the sound speed
inside the \hii~region. An isothermal jump conditions for the bow shock
reproduce fairly well the gas flow properties in simulations with ambient
gas density $\nH \gtrsim \ntwo$. Because the inner Bondi radius (for the
gas inside the \hii~region) is comparable but smaller than the I-front
radius, the accretion rate is well reproduced in our analytical model
assuming Bondi-type accretion from the ionized gas downstream of the
bow shock. The BH moves subsonically or transonically with respect to
the gas downstream of the bow-shock. In this regime, the accretion rate
increases with BH velocity because the density of the ionized gas inside
the \hii~region increases with increasing velocity reaching $n_{\rm H} \sim
\nH$ at $\vbh =2c_{\rm s,in}$.

\item Simulations of BH accreting from a high-density medium
  ($\nH=\nfive$) show steady accretion rate for all Mach numbers.
  However, at intermediate densities ($\nH=10^3$--$\nfour$) we find
  intermittent bursts of accretion rate in the Mach number range $2.5
  \lesssim \mach \lesssim \mach_R$. The oscillatory behavior of
  accretion rate is due to the development of instabilities in the
  thin shell in the upstream direction that cause its cyclic
  fragmentation and re-formation. In lower density regime, $\nH
  \lesssim \ntwo$, the post-shock density is lower than expected
  assuming isothermal shock jump conditions; as a result the dense
  shell is thicker and less prone to instabilities. For $\vbh > v_R$ the
  dense shell never forms, providing steady accretion rates.

\item Contrary to the case of radiation-regulated accretion onto
  non-moving BHs in which $\dot M \propto T_\infty^{5/2} \dot{M}_B
  \propto T_\infty$ (see Papers~I and II), the accretion rate onto
  supersonic BHs is independent of the temperature of the ambient
  medium (see Equation~(\ref{eq:res})). It follows that if $T_\infty
  <10^4$~K the accretion rate onto a BH moving with velocity $\vbh
  \approx 50~{\rm km~s^{-1}}$ is about $5(10^4~{\rm K}/T_{\infty})$
  times larger than the accretion rate at $\vbh=0$. Hence, the growth
  rate and the mean accretion luminosity of BHs moving supersonically
  can be significantly larger than that of non-moving BHs with the
  same mass and accreting from the same medium.
\end{itemize}

Our simulations are 2D, having assumed rotational
symmetry around the axis defined by the BH motion and a spherically
symmetric radiation field emitted near the BH. Below we discuss how
our results and conclusions might be sensitive to the inclusion of the
extra degree of freedom in more realistic three-dimensional
simulations and different assumptions on the feedback mechanism.

It is well known that Hoyle--Lyttleton accretion without radiation
feedback is prone to side to side ``flip-flop" instability
\citep[e.g.,][]{MatsudaIS:87, BlondinP:09, BlondinR:12} that cannot
be captured in our axis-symmetric simulations. This instability breaks
the axial symmetry of the flow downstream the BH where the gas,
gravitationally focused by the BHs, is shocked and accreted. The
resulting accretion of angular momentum produces the formation of
temporary accretion disks spinning in alternating directions, with a
burst of mass accretion when the direction of rotation is
flipped. This non-axisymmetric instability has been seen in numerous
hydrodynamic simulations of 2D planar accretion
\citep{Matsudaetal:91, ZarinelliWN:95, BenensohnLT:97, ShimaMABB:98,
PogorelovOM:00} but it has not been unambiguously identified in
three-dimensional simulations \citep{Ruffert:99}, possibly due to
insufficient spatial resolution. In our simulations, radiation feedback
may stabilize this instability for the cases, shown in Figure~3, when
a stable bow-shock forms upstream of the BH. In this regime, the
post-shock gas is transonic and the flow is better described by
Bondi-type accretion than Hoyle--Lyttleton. For this reason, and
because of the stabilizing effect of radiative heating
\citep{BlondinKF:90}, the flip-flop instability may not be important in
Hoyle--Lyttleton accretion when radiation feedback is included.  

An extra spatial degree of freedom is likely to play a role
in determining the stability of the thin cooling layer that forms
behind the bow shock. As seen in Figure~\ref{fig:den_d4}, probably because of the
axis-symmetric assumption in our simulations, the fastest growing mode in
the unstable shock is a ``dimple" forming along the axis of symmetry. Both
the instability growth rate and the shock break up conditions may
change given the additional degrees of freedom in three-dimensional
simulations. For instance, the spherical accretion shock instability
 found in supernova simulations
\citep[e.g.,][]{BlondinS:07} is less important in three-dimensional
simulations than in 2D ones.


In this paper, as in the previous papers in this series, we assume
that the radiation is emitted in a spherically symmetric manner near
the BH. However, it is possible that the formation of optically thick
structures near the BHs, such as an accretion disk, which are not
resolved in our simulations, may change the angular dependence of the
emitted radiation. For instance, \citet{ProgaSD:98} have studied
extensively the case of accretion on non-moving SMBHs in which UV
photons are emitted preferentially in the direction perpendicular to
the accretion disk. Here, we have not explored how such assumption
would affect our results for a moving BH. Neither we have explored the
feedback effect from a collimated wind or jet, that may also develop
in some phases of the BH accretion cycle.


Despite the aforementioned simplifying assumptions in our simulations,
our results are a substantial improvement with respect to the
Hoyle--Lyttleton model.  The dependence of the growth rate and
luminosity of BHs on their velocity with respect to the ambient gas is
qualitatively modified by radiation feedback. The results reported in
this paper may have significant repercussions in modeling and
understanding the growth of stellar BHs in the early universe, the
build up of an early X-ray background and provide important clues for
explaining ULXs 
\citep*{Krolik:81,Krolik:84,Krolik:04,RicottiO:04b,
  RicottiOG:05,Ricotti:07,StrohmayerM:09,VolonteriB:12}. We plan to
address some of these issues in forthcoming papers.

\acknowledgments 
The authors thank Richard Mushotzky, Chris Reynolds,
Eve Ostriker, Tiziana Di Matteo, and James Drake for constructive
comments and feedback. The numerical simulations presented in this
paper were performed using high-performance computing clusters
administered by the Center for Theory and Computation of the
Department of Astronomy at the University of Maryland (``yorp"), and
the Office of Information Technology at the University of Maryland
(``deepthought"). This research was supported by NASA grants NNX07AH10G
and NNX10AH10G and NSF CMMI1125285. We thank the anonymous referee for
constructive comments.

\bibliographystyle{apj}
\bibliography{pr12b}



\end{document}